\documentclass{aa}
\usepackage{graphicx}
\usepackage{lipsum}
\usepackage{multirow}

\usepackage{commath}
\usepackage[switch,columnwise]{lineno} 

\usepackage[super]{nth}
\usepackage{tikz}
\usetikzlibrary{shapes,arrows}
\usepackage{array}
\usepackage{subcaption}
\usepackage[varg]{txfonts}
\usepackage{amsmath}
\usepackage{booktabs}
\usepackage{placeins}

\usepackage{longtable}
\usepackage{geometry}
\usepackage{caption}
\usepackage{adjustbox}
\usepackage{float}
\usepackage{tabularx} 

\usepackage[varg]{txfonts}
\usepackage{commath}
\usepackage[super]{nth}

\usepackage{siunitx}
\definecolor{click_color}{RGB}{46,48,118}

\usepackage{mathrsfs}

\usepackage{graphicx}
\graphicspath{{figures/}}
\usepackage[varg]{txfonts}
\usepackage{commath}
\usepackage[super]{nth}

\usepackage{siunitx}
\DeclareSIUnit{\erg}{erg}
\DeclareSIUnit{\jansky}{Jy}
\DeclareSIUnit{\parsec}{pc}

\usepackage{soul}
\usepackage{xcolor}

\sethlcolor{green}

\newcommand{\rnew}{FRB~20240114A}

\usepackage{hyperref}
\hypersetup{
  colorlinks=true,
  linkcolor=click_color,
  citecolor=click_color,
  urlcolor=click_color
  }
\usepackage[capitalize]{cleveref}
\crefname{equation}{eq.}{eqs.}
\crefname{section}{Sect.}{Sects.}

\usepackage{txfonts}

\begin{document}

   \title{Modelling scintillation in Fast Radio Bursts}

\author{
    L. Nicotera\thanks{Email: lnicotera@mpifr-bonn.mpg.de}\inst{1}
    \and
    T. Sprenger\inst{1}
    \and
    O.Wucknitz\inst{1}
    \and
    L.G. Spitler\inst{1}
    \and 
    F. Eppel \inst{2}
    \and 
    P. Limaye \inst{3}
    }

\institute{
Max-Planck-Institut für Radioastronomie, Auf dem Hügel 69, 53121 Bonn, Germany
\and
Joint Institute for VLBI ERIC, Oude Hoogeveensedijk 4, 7991 PD Dwingeloo, The Netherlands
\and
LPC2E, OSUC, Univ Orléans, CNRS, CNES, Observatoire de Paris, F-45071 Orléans, France
}

\date{Received XXXX; accepted XXXX}


\abstract
{Fast radio bursts (FRBs) are bright, millisecond-duration radio transients observed at extragalactic distances. Propagation through inhomogeneities in the ionised medium produces scintillation, observed as frequency-dependent intensity variations in their spectra, with bright patches that are called scintles. The characteristic scintillation bandwidth, $\nu_s$, can probe scattering material along the line of sight. However, measuring this characteristic scale can be challenging when the available observing bandwidth of the burst spectra contains only a limited number of scintles.}
{We investigate the uncertainties associated with recovering $\nu_s$ from frequency autocorrelation functions (ACFs) under finite-bandwidth conditions. In particular, we examine the recovery of $\nu_s$ from individual ACFs and investigate whether averaging multiple independent ACFs can provide a more robust measurement.}
{Using simulations as a controlled framework, we reproduce observational conditions to investigate the modelling of scintillation in ACFs and the associated measurement uncertainties. We use these simulations to test the finite scintle error (FSE) prescription and assess its applicability to real observations. We then test the resulting modelling framework on Effelsberg Ultra Broad Band receiver observations of the repeating \rnew.}
{We find that, as the number of scintles decreases, individual ACFs develop prominent fluctuations, or "wiggles" at outer frequency lags, that deviate from the expected Lorentzian profile. This results in increased scatter in the recovered $\nu_s$ and a less reliable FSE prescription. Averaging multiple independent ACFs mitigates these effects and improves the recovery of $\nu_s$, particularly in the low-scintle regime.}
{Our results provide a practical framework for ACF modelling and uncertainty estimation in FRBs. For individual bursts, the FSE is effective with sufficient scintles; we find $\eta = 0.055$ more suitable for FRB analysis and update the FSE accordingly. For repeaters, instead, averaging multiple ACFs and bootstrap resampling improve the recovery and uncertainty estimation of $\nu_s$.}


\maketitle
\nolinenumbers
\section{Introduction}
\label{sec:Intro}
Fast radio bursts (FRBs) are energetic, millisecond-duration radio transients. Since their first discovery \citep{Lorimer2007}, thousands of FRBs continue to be detected in large-scale radio surveys (e.g. \citet{Abbott2026}). These sources are of particular interest due to their extragalactic origin, the coherent nature of their emission, and the still-uncertain physical mechanism responsible for their production \citep{Petroff_2019}. Among the proposed progenitor models, magnetars currently represent the most promising candidates. This interpretation is supported by the detection of FRB-like radio bursts associated with Galactic magnetars SGR 1935+2154 \citep{Andersen2019,Bochenek2020} and later 1E 1547.0-5408  \citep{lower2026transientnarrowbandradiobursts}.
Observational campaigns have successfully localised several FRBs to their host galaxies (e.g. \citet{Bannister2019}), extending to cosmological distances \citep{caleb2025fastradioburst3}. 

FRBs are commonly classified into repeating and apparently non-repeating (one-off) sources. The discovery of repeated bursts from FRB~20121102A \citep{Spitler_2016} established the existence of a repeating FRB class, suggesting that the observed diversity of FRB activity may arise from different progenitor channels. Beyond their origin, FRBs serve as powerful cosmic probes of the intervening baryonic matter, allowing constraints on the ionised, magnetised, and turbulent plasma along lines of sight \citep{Walters_2018, Macquart_2020}. 

During their journey across gigaparsec scales, FRBs traverse multiple plasma environments, including the interstellar medium (ISM) and circumgalactic medium (CGM) of the host galaxy, the intergalactic medium (IGM) and any intervening structures. They then pass through the Milky Way’s halo and ISM before they reach the detector via the interplanetary medium of the Solar System \citep{Macquart_2013, Ocker_2021}. FRB signals, therefore, interact with free electrons along the line of sight and as a consequence they undergo dispersion and multipath propagation. 

Small-scale inhomogeneities in the plasma density act as scattering screens, inducing multipath propagation that leads to temporal broadening, angular broadening, and interference patterns in the signal, known as diffractive scintillation \citep{Rickett1990, handbook}. Studying scintillation in FRBs provides a unique view on both the emission mechanism and the local plasma environment.

The paper is organized as follows. In Section~\ref{sec:pipeline}, we discuss scintillation in FRBs and we describe the scintillation analysis pipeline. In Section~\ref{sec:observational_data}, we  present the observational data, including the datasets used, the burst extraction procedure, and the data-processing steps. In Section~\ref{sec:simulated_data}, we describe the simulations, detailing the creation of synthetic data and the application of the analysis pipeline to the simulated datasets. In Section~\ref{sec:results}, we present the results and discuss their physical interpretation. In Section~\ref{sec:conclusion}, we summarize our conclusions.

\section{Scintillation in FRBs}
\label{sec:pipeline}
The propagation of FRBs through the ionized interstellar medium can imprint characteristic signatures on their emission, including temporal broadening and scintillation. Temporal broadening is characterised by the scattering timescale, $\tau_{\mathrm{s}}$, which increases at lower frequencies, $\tau_{\mathrm{s}} \propto \nu^{-\alpha}$, while scintillation produces frequency-dependent intensity variations, generating the so called "scintles”, with a characteristic decorrelation bandwidth, $\nu_s$, which grows with frequency, $\nu_s \propto \nu^{\alpha}$. For Gaussian plasma density fluctuations, the index $\alpha$ is typically close to 4.4, as expected for Kolmogorov turbulence. The parameters $\tau_{\mathrm{s}}$ and $\nu_s$ are inversely related \citep{Cordes_1998}; when the observed scintillation bandwidth and pulse broadening arise from the same scattering medium, they satisfy the relation $2\pi \, \nu_s \, \tau_d = C$, with $C$ in the range 0.5–2 \citep{Cordes_1998,Lambert_1999}. For FRBs, in the presence of multiple scattering regions, the scintillation bandwidth generally traces Galactic scattering, while the observed pulse broadening may instead arise from scattering beyond the Milky Way, most likely within the FRB host galaxy \citep{Masui_2015, Gajjar_2018, sammons2023twoscreenscatteringcraftfrbs}.

In practice, $\nu_s$ is commonly estimated by modelling the frequency autocorrelation function (ACF) of the burst spectrum. For a localised (thin) scattering screen that produces an approximately homogeneous scintillation pattern across the observing band, the ACF is well described by a Lorentzian function, whose half-width at half-maximum (HWHM) provides an estimate of $\nu_s$ \citep{Rickett1990}. When multiple scattering screens are present, additional Lorentzian components are required \citep{Pradeep2025}, e.g. \citet{nimmo2024magnetosphericoriginfastradio} used a double Lorentzian fit to best model the ACF of FRB 20221022A identifying one coherent scintillation scale associated with the Milky Way and one with the host galaxy. 

Hence, the scintillation bandwidth is a key observable for studying the plasma through which an FRB signal propagates. By measuring $\nu_s$, we can constrain the strength and location of the scattering screens along the line of sight, helping to characterise the turbulent ionized interstellar medium and the local environments of FRBs. Accurate measurements of $\nu_s$ are therefore essential for interpreting scintillation observations.

Another observable commonly used in scintillation studies is the modulation index, $m$, defined as the standard deviation of the intensity normalised by its mean, which provides a simple measure of the strength of the intensity fluctuations \citep{Rickett1990}. Thus, $m \sim 0$ corresponds to little or no intensity modulation, while $m \sim 1$ indicates strong fluctuations. For fully developed diffractive scintillation from a single scattering screen, $m$ is expected to approach unity. More complex propagation, including contributions from multiple scattering screens, can modify the intensity statistics and produce modulation indices either below or above unity.

While classical scintillation analysis is effective when scintillation features are densely and uniformly distributed across the observing band, FRB spectra often exhibit isolated scintles. This complicates the interpretation of the frequency ACF and can lead to poorly constrained estimates of $\nu_s$ when relying on single bursts. In the case of pulsars, where many pulses are typically available and can be averaged together, the classical approach provides robust measurements of $\nu_s$ \citep{Turner_2024}. In contrast, analyses of FRBs, particularly one-off sources, lack this averaging capability. Instead, analyses of repeating FRBs can provide more reliable constraints by combining multiple bursts, smoothing out spectral variations.

In this work, we investigate the challenges associated with modelling frequency ACFs in the presence of isolated scintles in FRB spectra. We aim to assess the uncertainties introduced by standard fitting procedures by comparing simulated and observational data, with the goal of quantifying error ranges and evaluating potential improvements in future analyses.

The observational dataset consists of \rnew, a repeating source that underwent a period of intense burst activity. We make use of the available bursts detected with the Effelsberg-100 m radio telescope \citep{Eppel_2025, Limaye_2026} to study its scintillation properties and to highlight the limitations of conventional modelling approaches when applied to band limited FRB spectra. In the following subsections, we summarize the methodology employed and the associated limitations of standard ACF-based techniques. 

\begin{figure}[ht]
\centering
\includegraphics[width=0.40\textwidth]{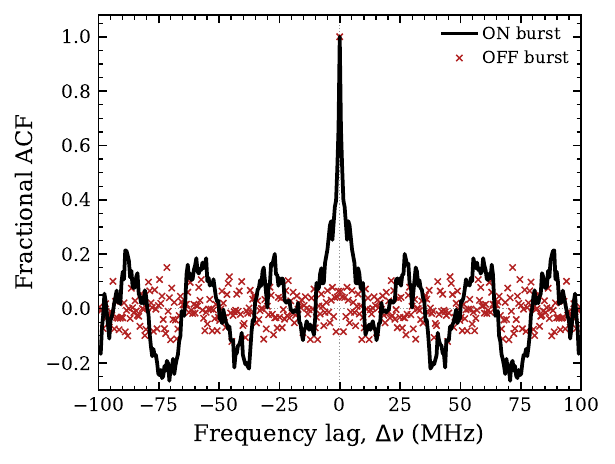}
\caption{Comparison of the frequency ACFs computed from the on- and off-burst regions of a representative \rnew\ burst observed with the Effelsberg UBB receiver (in Band~1, 1290--1939~MHz), after applying the bandpass normalization. The on-burst ACF shows the characteristic wiggles at larger frequency lags, while these structures are absent from the off-burst ACF, indicating that they are unlikely to originate from residual instrumental structure associated with the receiver backend.}
\label{fig:on_off_acfs}
\end{figure}

\subsection{Irregular scintillation structures}
\label{sec:wiggles}

FRB spectra that are band-limited may contain only a few bright scintles, leading to deviations from the expected ACF shape and characteristic "wiggles" across frequency lags. An example is shown in Fig~\ref{fig:on_off_acfs}, where these structures are clearly visible in the on-burst ACF. These features cause the ACF to deviate from the smooth Lorentzian shape expected for more uniform scintillation. 

For example, similar behaviour was reported by \citet{Levkov_2022} for FRB 20121102A, where oscillations and secondary peaks were observed in the frequency ACF. While scintillation alone is generally expected to produce a smooth decrease of the ACF with frequency lag, they suggested that the observed secondary maxima could arise from additional propagation effects, such as plasma or diffractive lensing, although an intrinsic origin related to the burst emission could not be ruled out. In particular, they interpreted the oscillatory structure as possible interference between two propagation paths, or images. However, scattering by plasma structures can produce a much larger number of images \citep{Simard_2018}. In our simulations, we also find oscillations and secondary peaks in the ACF, despite not including a specific two-image configuration. This suggests that such features are not unique to two-path interference and can also arise from other effects, such as the finite-bandwidth of the spectra. Similar structures are also observed in the ACFs of \rnew.

These considerations motivate several important questions. How much of the irregular structure in the ACF should be considered physically meaningful and included in the scintillation modelling? Or should the analysis focus on the central region, where the decorrelation bandwidth is most clearly defined and the irregular structure is less prominent?

In the next subsection, we describe the fitting strategy adopted in our pipeline. Starting from the standard Lorentzian model, we introduce several modifications to improve the reliability of the fit in the presence of sparse scintillation patterns and irregular ACF features.

\begin{figure*}[t]
\centering
\includegraphics[width=0.83\textwidth]{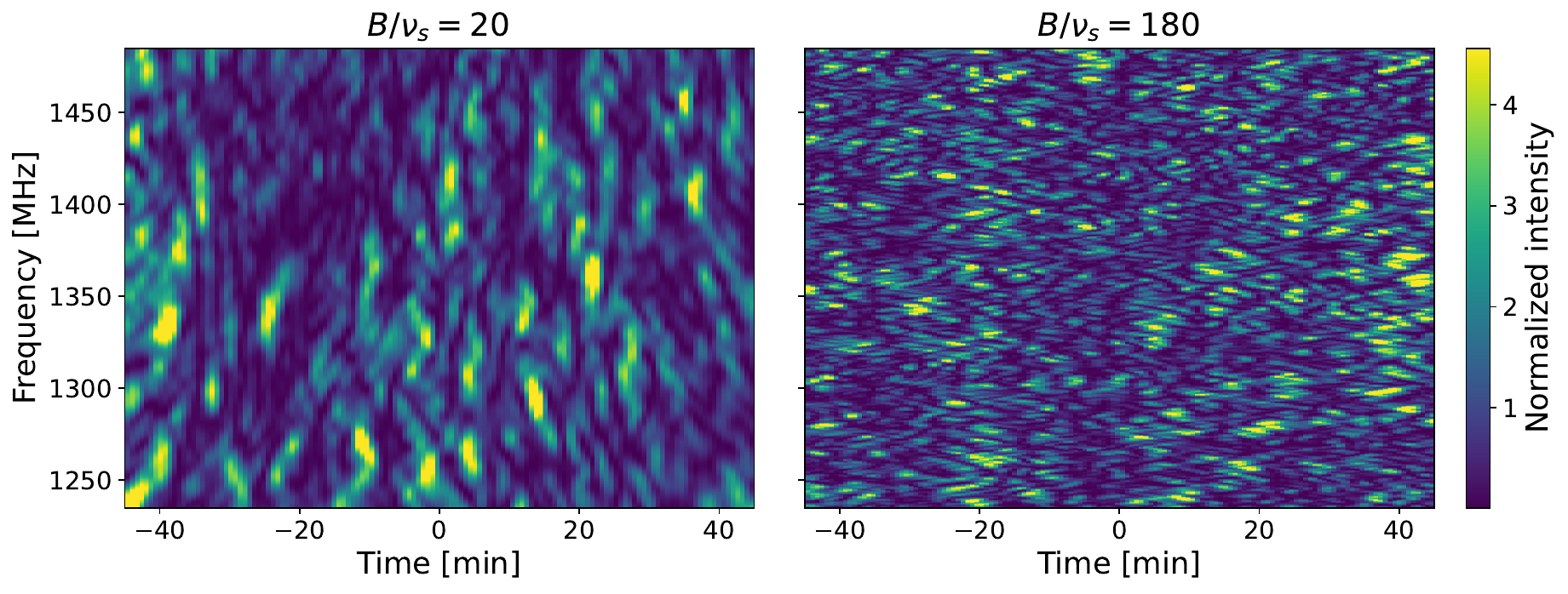}
\caption{
Simulated two-dimensional scintillation fields for two different values of the ratio between the observing bandwidth $B$ and the injected scintillation decorrelation bandwidth $\nu_s$. Left: $B/\nu_s = 20$, resulting in more isolated scintles across the band. Right: $B/\nu_s = 180$, producing a much denser scintillation pattern with a larger number of independent scintles.
}
\label{fig:2d_scintillation_fields}
\end{figure*}

\subsection{FRB scintillation analysis pipeline}
\label{sec:FRB_pipeline}

We developed a pipeline to estimate the scintillation bandwidth by modelling the spectral ACF of the burst spectrum with a classical Lorentzian profile. The aim of the procedure is to obtain a stable estimate of $\nu_s$, while minimizing the impact of noise, residual bandpass structure, and large-lag fluctuations that can dominate the fit, when the spectrum contains isolated scintles. 
Before computing the ACF, frequency channels contaminated by radio frequency interference (RFI) are excluded from the analysis. We then perform a bandpass correction using the off-burst data. For each unmasked frequency channel, we calculate the mean and standard deviation of the off-burst intensity over time. The on-burst dynamic spectrum is normalised by subtracting the corresponding off-burst mean and dividing by its standard deviation. Finally, the normalised on-burst data are averaged over time to obtain the one-dimensional spectrum, $I(\nu)$, used for the ACF analysis. Following the approach of \citet{Pradeep2025}, we define the fractional fluctuations of $I(\nu)$ as

\begin{equation}
x(\nu_j) =
\frac{I(\nu_j)-\langle I \rangle}{\langle I \rangle},
\label{eq:fractional_fluct}
\end{equation}
where $\langle I \rangle$ is the mean intensity over the unmasked frequency channels. This normalization removes the overall intensity scale and isolates the relative spectral modulations associated with scintillation. We then compute the spectral ACF from the fractional spectral fluctuations as a function of frequency lag $\Delta\nu$, using all valid pairs of frequency channels $N_{\rm pair}(\Delta\nu)$, while preserving the original channel spacing after RFI excision, as

\begin{equation}
{\rm ACF}(\Delta\nu) = 
\frac{1}{N_{\rm pair}(\Delta\nu)}\sum_{j}x(\nu_j)\,x(\nu_j+\Delta\nu).
\label{eq:acf}
\end{equation}

As a check for residual instrumental structure, in Fig~\ref{fig:on_off_acfs} we compare the ACFs computed from the on- and off-burst regions of one representative \rnew\ burst from our Effelsberg-100 m Ultra-Broad-Band (UBB) receiver observations, after applying the bandpass normalization. The characteristic structures observed in the on-burst ACF are not present in the off-burst ACF, indicating that they are unlikely to originate from instrumental effects associated with the polyphase filterbank of the receiver backend. We therefore exclude this instrumental origin as the primary cause of the observed ACF structure and proceed with the scintillation analysis.

Measurement noise contributes primarily as a sharp spike at zero lag in the ACF. To avoid biasing the fit, we account for this effect by excluding a small range around $\Delta\nu=0$ from the model optimization, specifically the zero lag and the first adjacent lag on either side. The remaining lags are then fitted with the Lorentzian model defined as: 
\begin{equation}
L(\Delta\nu) = \frac{A}{1+(\Delta\nu/\nu_s)^2}+C,
\label{eq:lorentzian}
\end{equation}
where $A$ is the amplitude, $\nu_s$ is the scintillation bandwidth, and $C$ is a constant offset included to account for possible residual deviations of the ACF baseline from zero.
In this analysis, we assume a single scattering screen, fitting one Lorentzian component to measure a single scintillation frequency scale. To reduce the impact of noise at large frequency lags, we apply a triangular weighting function during the fit. The weighting scale is determined from an initial estimate of the scintillation bandwidth obtained directly from the measured ACF using a half-maximum estimator. This allows the fit to focus on the central region. The weight assigned to each frequency lag is defined as

\begin{equation}
w(\Delta\nu) =
\max\left(0,\,1-\frac{|\Delta\nu|}{W}\right),
\label{eq:triangular}
\end{equation}
where the characteristic weighting scale is $W = 8\nu_{s,0}$, and $\nu_{s,0}$ is the initial estimate of the scintillation bandwidth. The weighting therefore decreases linearly with increasing $|\Delta\nu|$, giving greater weight to the central part of the ACF and progressively downweighting points at larger frequency lags.
The Lorentzian parameters are obtained using a bounded least-squares fit with a robust Cauchy loss function. 

As an additional consistency check, we also analyze the modulation power spectrum of the mean-subtracted burst spectrum, obtained as the Fourier transform of the frequency ACF. Under the assumption of a Lorentzian ACF, the corresponding power spectrum is expected to follow a two-sided exponential form \citep{Reardon_2019}, providing an independent estimate of $\nu_s$ (see \cref{sec:appendix_simulation} for a derivation):
\begin{equation}
P(f) =
A_{\rm ps}\exp\!\left(-2\pi\nu_{\rm s}\,|f|\right)+C_{\rm ps},
\label{eq:power_spec}
\end{equation}
here $f$ is the Fourier-conjugate variable to frequency and therefore it represents the time lag, $A_{\rm ps}$ is a normalization constant and $C_{\rm ps}$ accounts for a possible residual offset in the power spectrum.

Although the power-spectrum estimate is not used in the analysis presented here, this step is included in the pipeline for completeness and provides an additional consistency check on the agreement of the independently recovered values of $\nu_s$ with those obtained from the ACF. 

Another parameter that we calculate is the modulation index, conventionally defined as $m=\sigma_I/\langle I\rangle$, where $\langle I\rangle$ is the mean intensity and $\sigma_I$ is the standard deviation of the intensity across the considered frequency range. In our analysis, following \citet{Pradeep2025}, We estimate the modulation index from the zero-lag value of the Lorentzian fitted to the ACF as
\begin{equation}
m = \sqrt{A+C}.
\label{eq:modulation_index}
\end{equation}

Finally, we estimate the uncertainty on $\nu_s$, following the finite scintile error (FSE) approach of \citep{cordes1986, brown2025} as 
\begin{equation}
\sigma_{\nu_{\rm s}}
= \frac{\nu_{\rm s}} {2\ln 2\,\sqrt{N_{\rm d}}},
\label{eq:sigma_fse}
\end{equation}
where the effective number of independent scintles across the observing band is

\begin{equation}
N_{\rm d} = 1 + \eta\frac{B}{\nu_{\rm s}}
\label{eq:number_scint}
\end{equation}
\citep{Cordes_1998}. Here, $B$ is the usable receiver bandwidth and $\eta$ is a filling factor. 
A scintle filling factor of $\eta\sim0.1$--$0.3$ is typically assumed in the literature \citep{cordes2010measurementmodelprecisionpulsar}. 

We note that the uncertainties discussed throughout this work are systematic, while the contribution from thermal noise is not included.

\begin{figure*}[ht]
\centering

\begin{minipage}[c]{0.31\textwidth}
    \centering
    \includegraphics[width=\textwidth]{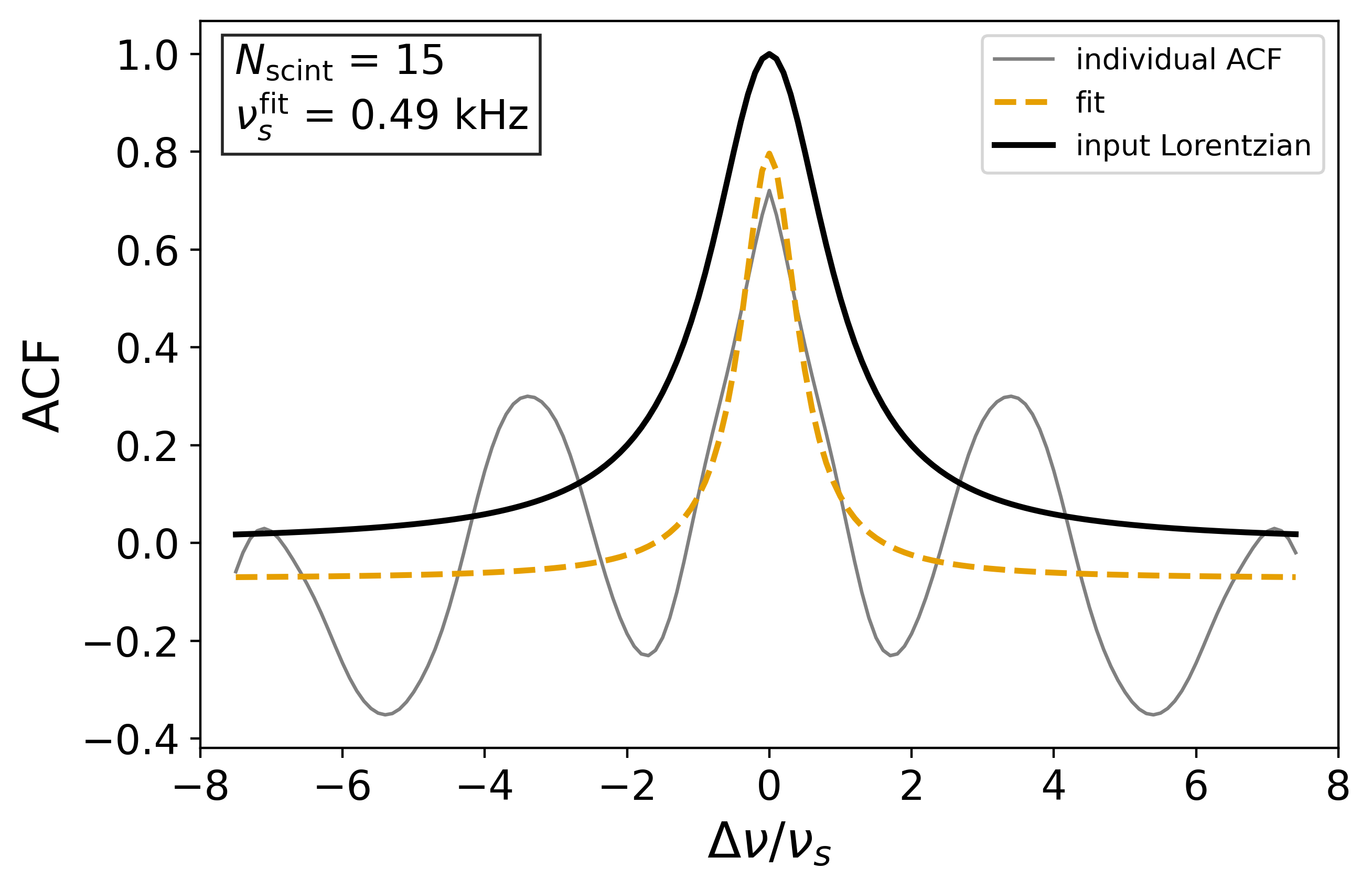}
\end{minipage}
\hfill
\begin{minipage}[c]{0.31\textwidth}
    \centering
    \includegraphics[width=\textwidth]{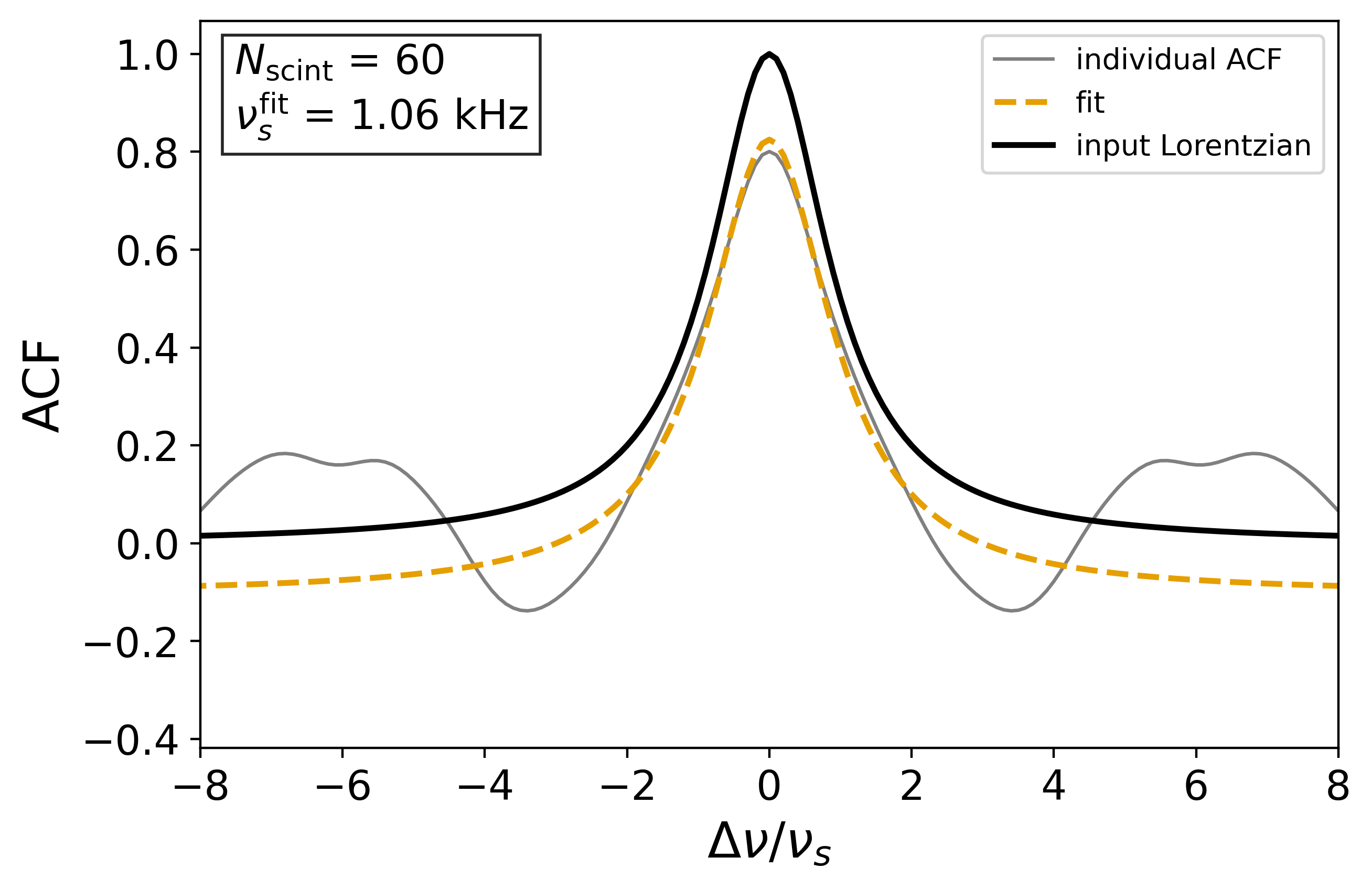}
\end{minipage}
\hfill
\begin{minipage}[c]{0.31\textwidth}
    \centering
    \includegraphics[width=\textwidth]{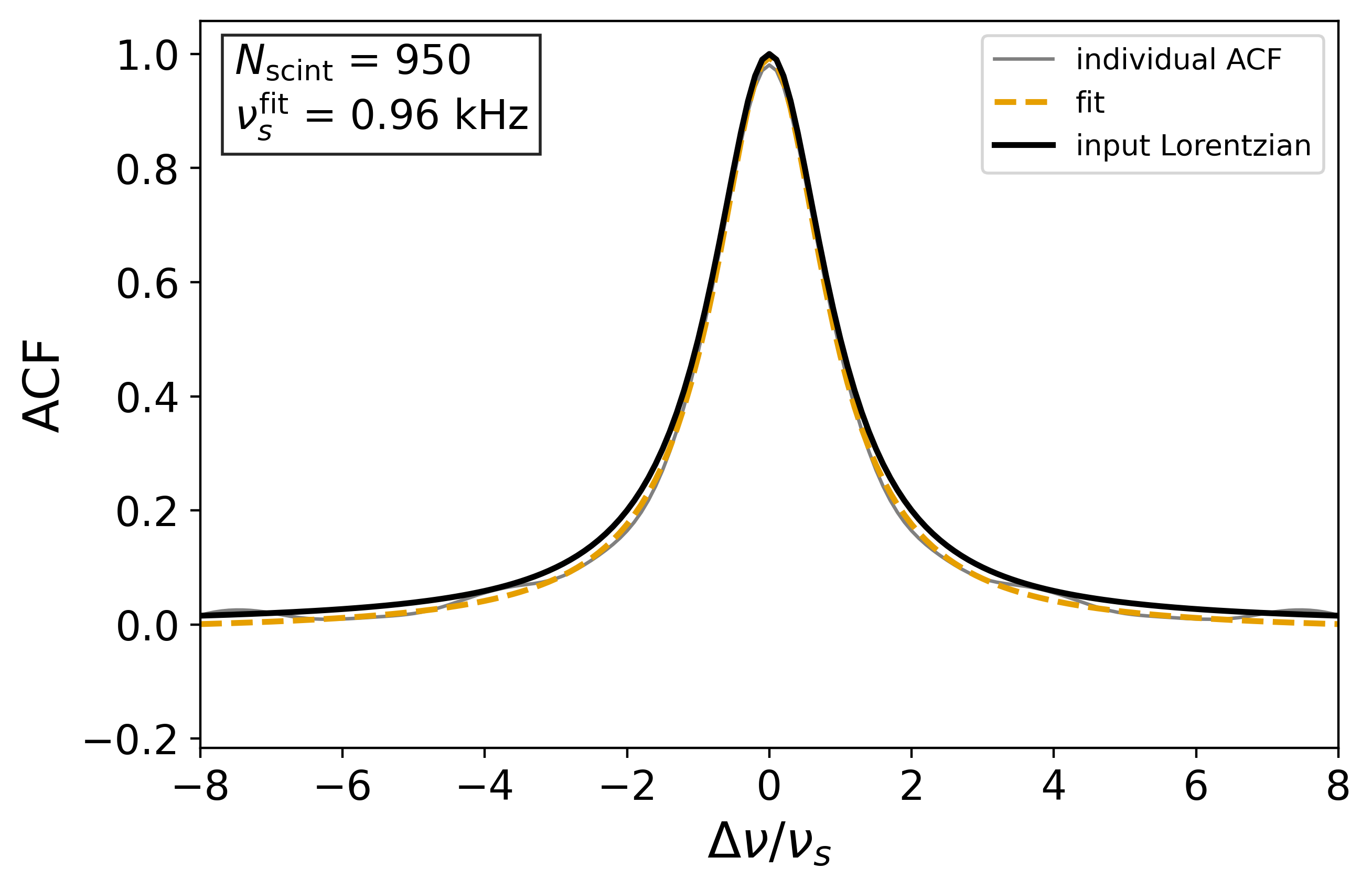}
\end{minipage}

\vspace{0.1cm}

\begin{minipage}[c]{0.34\textwidth}
    \centering
    \includegraphics[width=\textwidth]{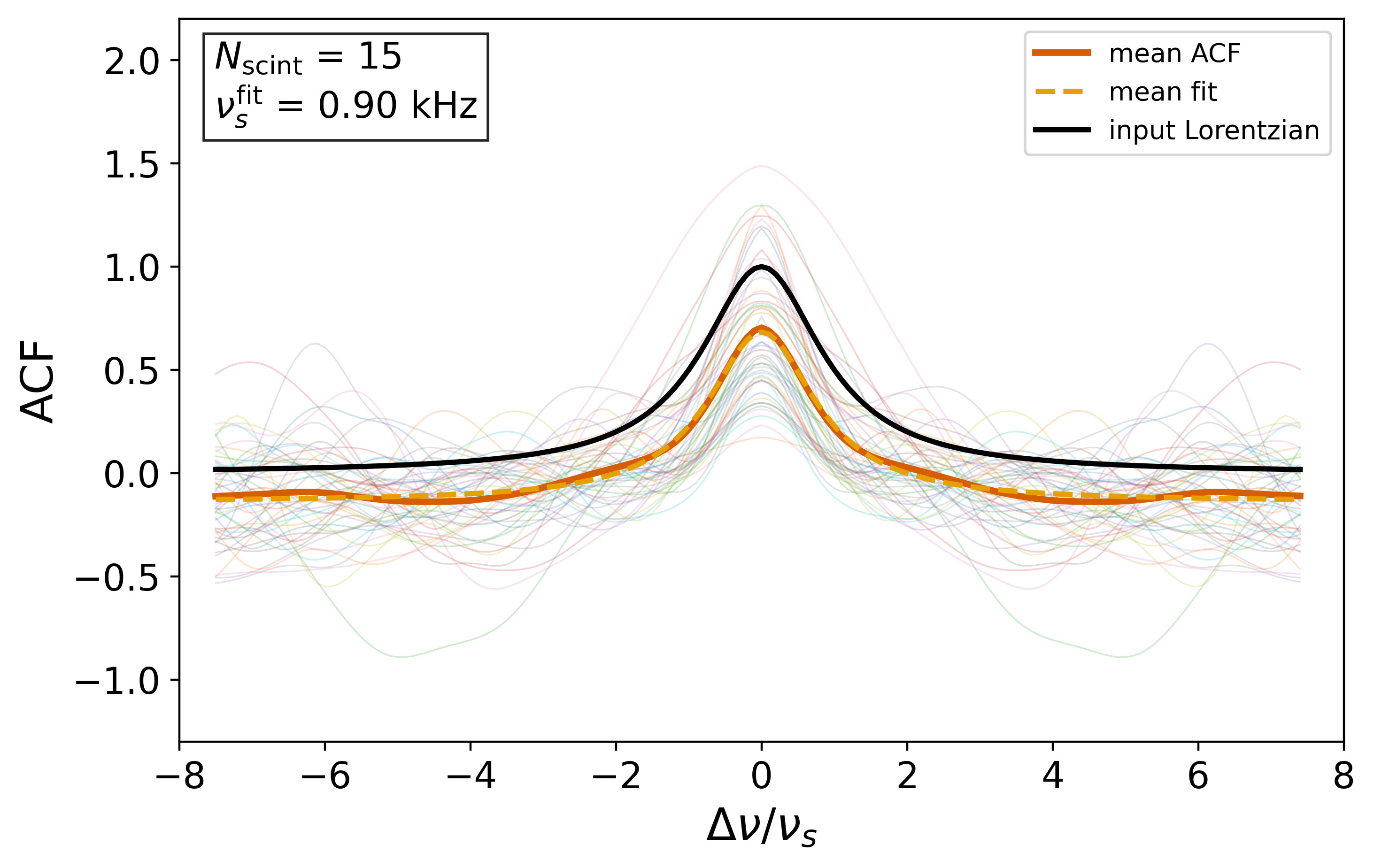}
\end{minipage}
\hspace{0.02\textwidth}
\begin{minipage}[c]{0.34\textwidth}
    \centering
    \includegraphics[width=\textwidth]{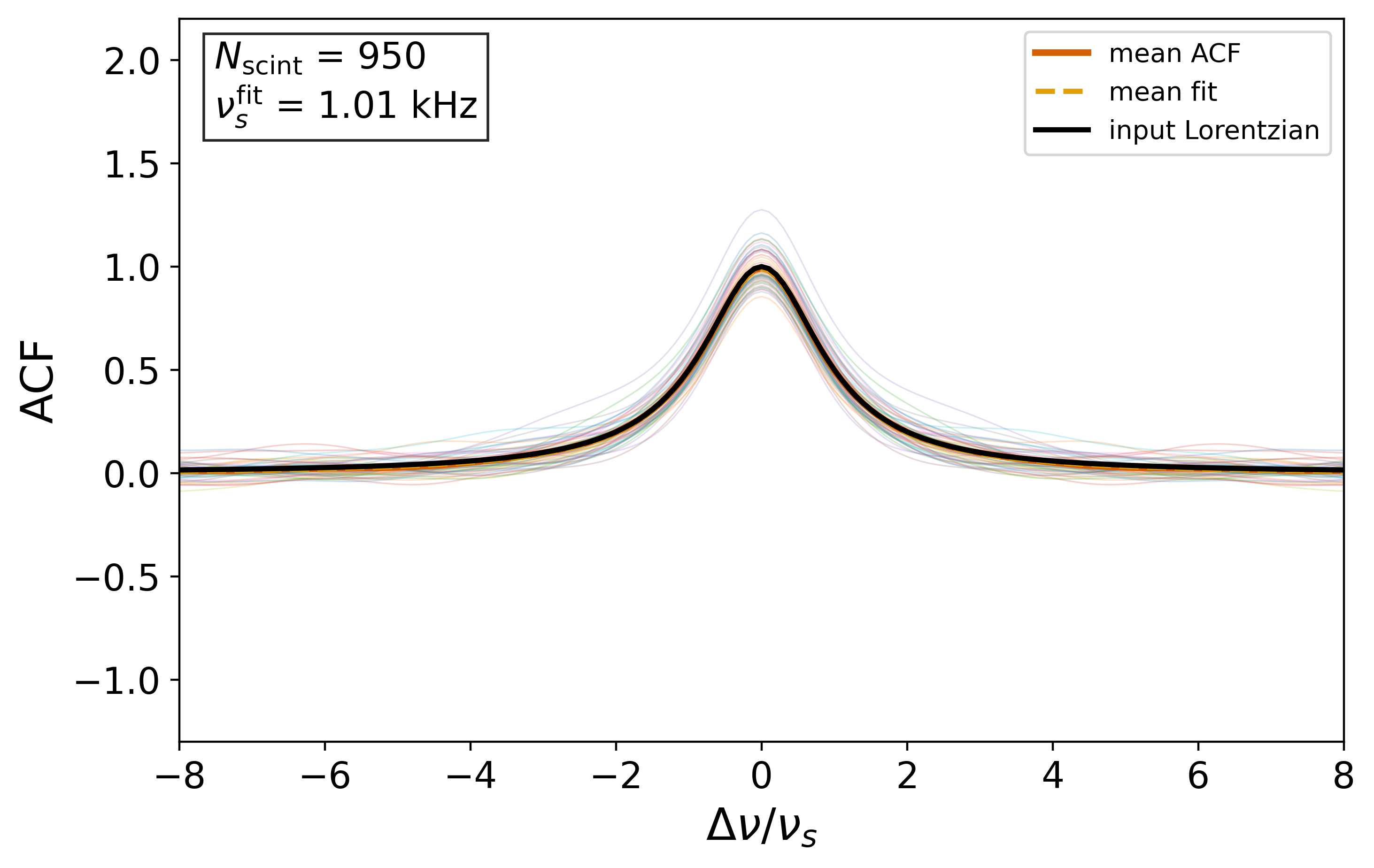}
\end{minipage}

\caption{Top: Examples of individual simulated ACF realizations (grey) for different scintle numbers, compared with the theoretical Lorentzian expectation (black). The yellow dashed line represents the fit obtained with our analysis pipeline. As the number of scintles decreases (right to left), the ACF becomes increasingly noisy and exhibits larger deviations from the ideal profile. Bottom: Comparison between the theoretical Lorentzian ACF (black) and the mean ACF (red) obtained by averaging 50 independent simulated ACFs, shown in the background. The dashed orange curve represents the Lorentzian fit recovered by our developed pipeline. 
In all simulations, the input scintillation bandwidth is $\nu_{s,\mathrm{input}}=1~\mathrm{kHz}$, while the boxes report the fitted $\nu_s$ for both the individual realizations (top panels) and the mean ACFs (bottom panels).}
\label{fig:simulations}

\end{figure*}

\section{Observational data}
\label{sec:observational_data}
\label{sec:r147}
\rnew\ is a recently discovered repeating fast radio burst source first reported by the CHIME/FRB Collaboration in early 2024 \citep{Shin2024}. Shortly after its detection, the source entered an exceptionally active bursting phase, rapidly becoming one of the most hyperactive repeaters observed to date and triggering extensive follow-up campaigns across multiple radio observatories, including FAST, MeerKAT, GMRT, Parkes, and Effelsberg (e.g. \citet{Panda_2025, uttarkar2026fastradioburstcyclone, Limaye_2026}). 

In this work, we analyse data observed with the Effelsberg UBB receiver system \footnote{\footnotesize See the Effelsberg receiver documentation: \url{https://eff100mwiki.mpifr-bonn.mpg.de/doku.php?id=information_for_astronomers:rx_list}.}, covering approximately 1.3--6.0~GHz in five subbands and recorded with the Effelsberg Direct Digitisation (EDD) backend \citep{EDDpaper}. 
Data detection is explained in detail in \citet{Eppel_2025} and \citet{Limaye_2026}. We focus on the most active observing epoch, corresponding to 23 May 2024 (MJD 60453), which contains several hundred bursts. These exhibit complex spectro-temporal morphologies, and strong frequency-dependent activity \citep{Limaye_2026}, with a significant fraction showing band-limited structure. 

To build our own high-quality sample for scintillation analysis, we performed a single-pulse search on the search-mode data using the BLInd Single Pulse Search (BLISS) pipeline \citep{Bause_2026}, based on TransientX \citep{men2024transientxhighperformancesingle} and PRESTO \citep{2001PhDT.......123R}. We then selected the highest signal-to-noise and most broadband bursts for further analysis.
The data have a frequency resolution of 500 kHz, just sufficient to resolve the scintillation bandwidth predicted for this line of sight from UBB Band 1 onwards. For \rnew, the NE2025 Galactic electron-density model \citep{Ocker_2026} predicts a scintillation bandwidth of approximately 0.52 MHz at a central frequency of 1 GHz. Previous observations, however, have reported different scintillation behaviour across frequency. Using MeerKAT UHF observations, \citet{tian2024detectionlocalisationhighlyactive} measured a scintillation bandwidth of $\nu_s = 4.36 \pm 0.11$ MHz at 847 MHz, substantially larger than the Galactic prediction. 
In contrast, \citet{Shin_2026} found no convincing evidence for astrophysical scintillation in the CHIME range of 400--800 MHz band, although they noted that the non-detection could be influenced by systematics, self-noise, or decoherence. Together, these results motivate the importance of measurements over a broad frequency coverage, like the one by the Effelsberg UBB observations. We apply our scintillation analysis pipeline to the selected bursts across the five UBB subbands to independently measure the scintillation properties with our fitting procedure. While the ACFs of individual bursts often exhibit substantial variability due to band-limited and irregular spectral structure, by averaging multiple ACFs we obtain a more reliable estimate of the scintillation bandwidth. This is a key advantage of repeating FRBs for scintillation studies. 

\section{Simulated data}
\label{sec:simulated_data}
Before investigating the impact of scintillation on individual FRB spectra in the next subsection, we first illustrate the underlying scintillation field from which these spectra are obtained. Fig~\ref{fig:2d_scintillation_fields} shows simulated two-dimensional scintillation fields with different scintle densities. In particular, as the observing bandwidth ($B$) increases and becomes much larger than the injected scintillation decorrelation bandwidth $(\nu_{s})$, the scintillation pattern evolves from a sparse field containing isolated bright scintles (left) to a more densely populated field (right). Since an observed FRB samples the scintillation field at a single instant in time, each burst spectrum corresponds to a one-dimensional slice through one of these fields. This visualisation provides an intuitive picture of how the underlying pattern gives rise to the spectral structures observed in individual FRBs. We do not describe the generation of these two-dimensional scintillation fields here; instead, the following subsection details the simulations used to generate the individual FRB spectra and their corresponding ACFs, which form the basis of the analysis presented in this work. Both the two-dimensional time-frequency and one-dimensional frequency-only simulation implementations are publicly available. \footnote{https://github.com/SprengerT/ScintillationMaker}.

\subsection{Simulation design}
\label{sec:simulation_design}
To investigate how limited scintillation patterns affect the morphology of the frequency ACF, we developed a simulation framework that generates synthetic scintillation spectra with a controlled scintillation bandwidth. We consider the simplified case of a single effective two-dimensional scattering screen wherein the resulting ACF is classically expected to follow the single Lorentzian profile with scintillation bandwidth $\nu_s$. Our simulations assume a constant scintillation bandwidth across the observing band. While this approximation is expected to be valid for a relatively narrow bandwidth, frequency-dependent scintillation may affect the ACF morphology. We plan to investigate this effect in future work.

Details on the simulation method are described in Appendix~\ref{sec:appendix_simulation}. The simulation generates a complex electric field spectrum $E(\nu)$ over a total bandwidth $B$, sampled with $N_\nu$ frequency channels. The observed intensity spectrum is then obtained from $I(\nu)=|E(\nu)|^2$ and normalised by its mean intensity in order to isolate the scintillation fluctuations. The characteristic scintillation bandwidth is introduced as an input parameter, hereafter denoted as $\nu_{s,\rm input}$, and sets the typical frequency scale of the spectral structures. Another input parameter is the number of images, $N_{\rm im}$, which represents the number of scattered wave components that are summed to produce a single realisation of the scintillation intensity. In the simulations presented here, the scintillation bandwidth is sampled by 10 frequency channels. The full synthetic spectrum contains 1000 scintles, corresponding to a total of 10000 frequency channels, and is generated from the interference of 10000 scattered image components.
To study finite-scintle effects, we progressively truncate the simulated spectrum into subsets with different bandwidths, corresponding to different numbers of scintles across the band, given by $N_{\rm scint}\sim B/\nu_{s,\rm input}$. We denote the intensity spectrum within each of these subsets as $I_{\rm sub}$ and use the corresponding values of $N_{\rm scint}$ to indicate our simulated cases. As the frequency range is progressively truncated, both the available bandwidth and the number of frequency channels decrease, while the frequency resolution remains fixed. For each $I_{\rm sub}$, we calculate the fractional intensity fluctuations following Equation~\ref{eq:fractional_fluct}, using the mean intensity within the corresponding frequency range. The simulation then computes the frequency ACF and compares it with the theoretical Lorentzian prediction.
Fig~\ref{fig:simulations} illustrates that, as the number of scintles decreases (right to left), the recovered ACF exhibits increasingly irregular structure across frequency lags, resulting in progressively larger deviations from the ideal Lorentzian profile. 

To quantify these effects, we generate 1500 independent spectra for each $N_{\rm scint}$; these are processed through the scintillation analysis pipeline described in Section~\ref{sec:FRB_pipeline}, following the methodology that we present in Section~\ref{sec:uncertainties}. This simulation therefore provides a controlled framework for evaluating under which conditions the recovered ACF converges to the theoretical expectation and for assessing how irregular scintillation features affect the stability and interpretation of Lorentzian ACF fitting in FRB observations.

\subsection{Assessment of the uncertainty}
\label{sec:uncertainties}
We now investigate the uncertainty associated with the recovery of $\nu_s$. For each $N_{\rm scint}$ case, we randomly select 50 realizations from the simulated ensemble and calculate the standard deviation of the recovered $\nu_s$ values, which we use as an empirical estimate of the uncertainty on an individual ACF measurement. We choose 50 as a representative sample size for the number of bursts that may be available in real observations. To assess the stability of this estimate, we repeat the procedure 1000 times for each $N_{\rm scint}$ case, creating a new random sample of 50 realizations each time and calculating the corresponding $\sigma_{\nu_s}$. We then compare these empirical uncertainties with the FSE predicted by Equation~\ref{eq:sigma_fse}, using the known spectral bandwidth and input scintillation bandwidth of the simulations. This provides a controlled test of whether the FSE correctly describes the uncertainty of an individual band-limited spectrum and how sensitive this comparison is to the particular set of realizations considered. We use the standard deviation of the recovered $\nu_s$ values across 50 realizations rather than the formal uncertainty returned by the ACF fit, as the latter can substantially underestimate true measurement uncertainties. The resulting empirical scatter therefore provides a more realistic basis for comparison with the FSE prediction.

A similar procedure is used to recover an empirical value of the filling factor, $\eta$. For each $N_{\rm scint}$ case and each of the 1000 random selections, the measured $\sigma_{\nu_s}$ is used to infer $\eta$ by inverting the finite-scintle relations in Equations~\ref{eq:sigma_fse} and~\ref{eq:number_scint}. This results in a distribution of 1000 inferred $\eta$ values for each $N_{\rm scint}$ case, which we characterize using its median and the 15th--90th percentile interval.

\subsection{ACF averaging for repeating FRBs}
\label{sec:acf_averaging}
We further investigate the effect of averaging multiple independent ACFs on the recovery of the scintillation bandwidth, and how this depends on the number of ACFs included in the average. For each $N_{\rm scint}$ case, the 1500 simulated spectra are divided into groups of different sizes, $N_{\rm group}$, and their corresponding ACFs are averaged before applying the Lorentzian fitting procedure. By varying $N_{\rm group}$, we assess whether increasing the number of averaged ACFs leads to a more stable recovery of $\nu_s$. This is particularly relevant for repeating FRBs, where multiple independent burst spectra may be available and can be combined to obtain a more robust estimate of the characteristic scintillation bandwidth.

\section{Results and discussion}
\label{sec:results}
In this section, we present the results of modelling the scintillation properties in both the simulated and observational data. We first identify how the finite number of scintles affects the recovery of the scintillation bandwidth, $\nu_s$, and its associated uncertainty. We then use these results to evaluate the finite scintle error prescription and constrain the filling factor $\eta$, to see under which conditions we could  subsequently apply it to the individual burst measurements of \rnew. Finally, we examine how averaging different numbers of independent ACFs affects the stability of the recovered $\nu_s$. 

\subsection{Individual ACF measurements and finite-scintle uncertainties}
\label{sec:individual_acfs}

\subsubsection{Simulation}
\label{sec:simulation_results}

To study the effect of finite bandwidth on individual ACF modelling, we analyse each simulated spectrum independently. For each subband, corresponding to a different $N_{\rm scint}$, we compute the ACF of the subbanded intensity spectrum, as described in Section~\ref{sec:simulation_design}, and apply the fitting procedure of our pipeline, described in Section~\ref{sec:FRB_pipeline} to recover the scintillation bandwidth. 

In the top panels of Fig~\ref{fig:simulations}, we show the results for three different  $N_{\rm scint}$; as the bandwidth is reduced and fewer scintles are included, the ACF becomes less smooth and increasingly deviates from the ideal Lorentzian shape. In particular, for the case of 950 scintles (top right), corresponding to nearly the full simulated bandwidth, the ACF from an individual realization is generally in good agreement with the theoretical Lorentzian model. Similar behaviour is observed for most of the displayed 50 realizations in the lower right panel plot of the same Figure. However, this represents an ideal scenario and differs from what we observe in our data. 

At lower numbers of scintles, such as the 60- and 15-scintle cases (see also top left and top central panels of Fig~\ref{fig:simulations}), the ACFs become increasingly irregular and more closely resemble the observational data, with the characteristic wiggles becoming particularly prominent for 15 scintles. This makes it more likely that the fitting procedure will recover a scintillation bandwidth that deviates from the true value, representing an important limitation when dealing with the finite bandwidth available for real FRBs. The triangular weighting (Equation~\ref{eq:triangular}) used in the fitting procedure helps to reduce the importance of the outer lags and gives more weight to the central part of the ACF, where the decorrelation scale is measured. However, the fitting remains unreliable for particularly noisy ACFs. 

We also explored an alternative normalization in which the fractional intensity fluctuations are defined as $(I_{\rm sub}-\langle I_{\rm full}\rangle)/\langle I_{\rm full}\rangle$, where the mean is computed from the full simulated spectrum rather than from the subbanded spectrum. With this normalization, the individual ACFs obtained for different $N_{\rm scint}$ become considerably more similar to one another, reducing the diversity of features seen across the different subband sizes in Fig~\ref{fig:simulations}. This demonstrates that the offset originates from the use of the subband mean in the normalization procedure. Nevertheless, we adopt the subband-based normalization throughout, as it better reflects real observations, where the mean of the full underlying spectrum is not available.

In summary, our simulations demonstrate that finite bandwidth limits the reliability of individual ACF fittings, with fewer scintles increasing deviations from the true scintillation model.

\subsubsection{Testing the finite scintle error prescription}
\label{sec:test_fse}

\begin{figure} [H]
\centering

\includegraphics[width=0.84\linewidth]{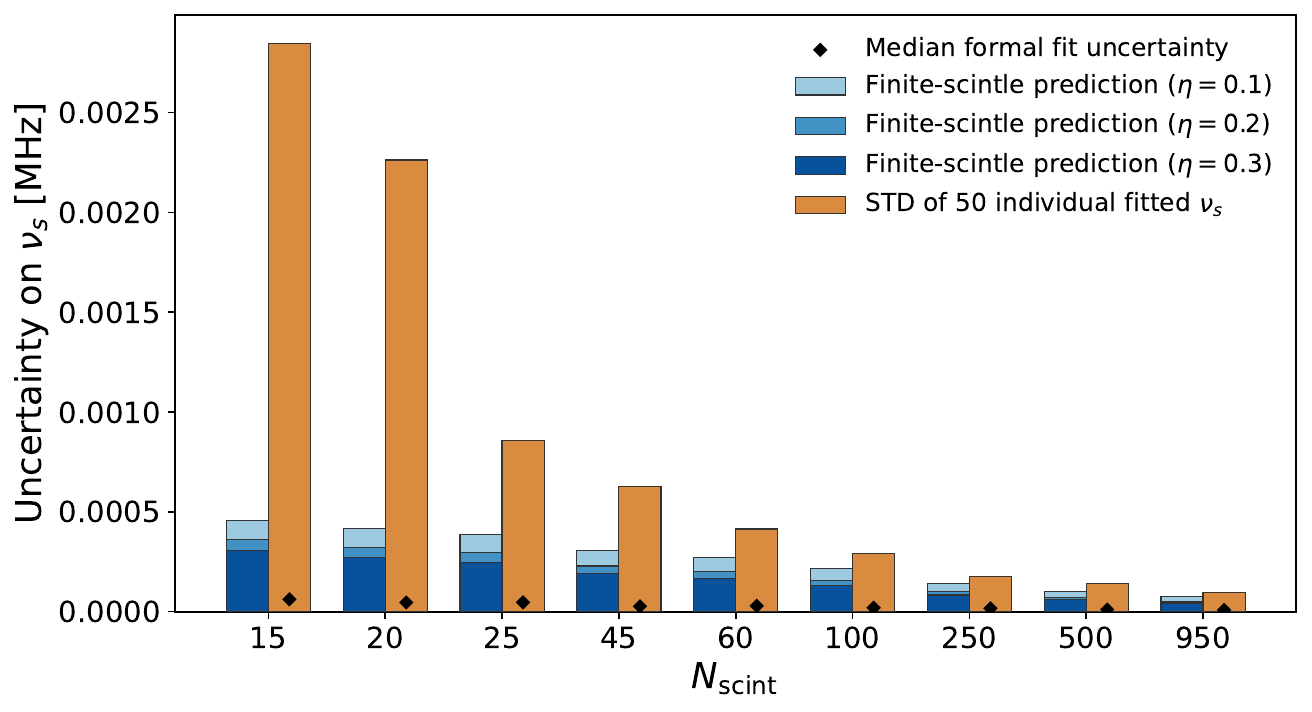}
\caption{
Comparison between the empirical uncertainty on the scintillation bandwidth and the FSE. For each simulated scintle case, the orange bars show the standard deviation of the recovered scintillation bandwidths across 50 randomly selected realizations. The FSE predictions (Equation~\ref{eq:sigma_fse}) are evaluated using $\nu_{\rm s,input}$ for three different values of the filling factor, $\eta=0.1$, $0.2$, and $0.3$, and are shown using different shades of blue. The black diamonds represent the median of the formal uncertainties from the individual ACF fits, highlighting that the fitting uncertainties are substantially smaller than the empirical scatter and therefore underestimate the true measurement uncertainty, particularly at low $N_{\rm scint}$. The same comparison is repeated for 1000 random selections of 50 realizations, with qualitatively similar behaviour. }
\label{fig:fse_validation}
\end{figure}

\begin{figure} [H]
\centering
\includegraphics[width= 0.78\linewidth]{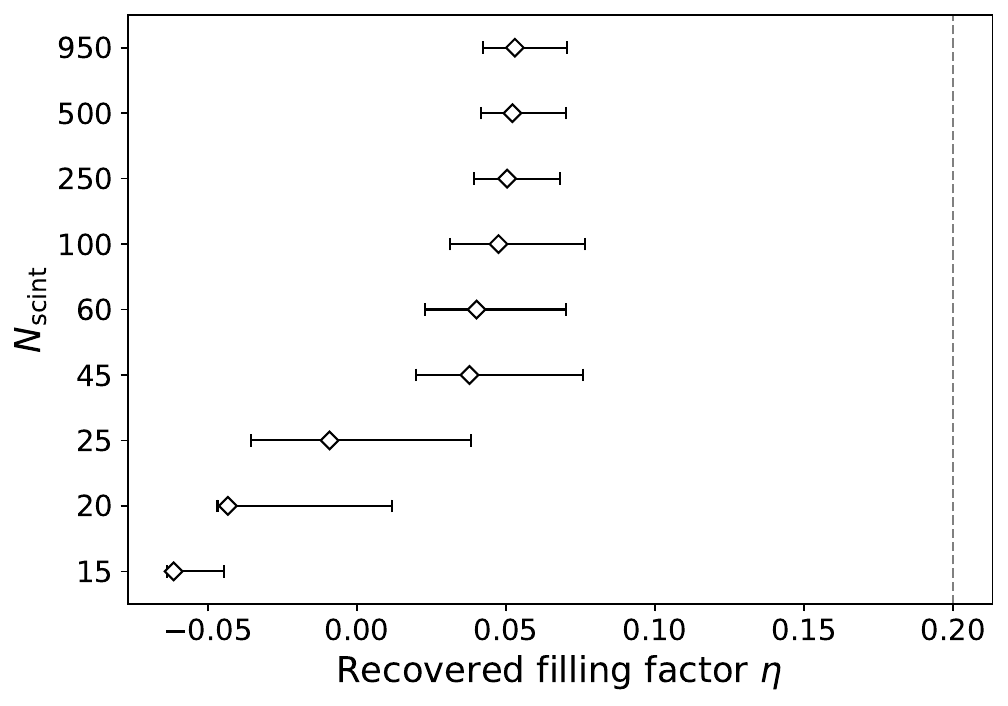}
\caption{Recovery of the effective filling factor, $\eta$, for the different simulated scintle cases. For each $N_{\rm scint}$, $\eta$ is inferred from the measured scatter of the scintillation bandwidths recovered from the simulated spectra and using the known $\nu_{\rm s,input}$ in the FSE. The analysis is repeated for 1000 random selections of 50 realizations from the full simulated ensemble. White diamonds show the median recovered $\eta$, while the horizontal error bars indicate the 15th--90th percentile range of the resulting distributions. For the $N_{\rm scint}\leq 25$ cases, the inversion yields $\eta<0$, which has no physical interpretation. The vertical dashed line marks the reference value $\eta=0.2$.}
\label{fig:eta_recovery}
\end{figure}

We now use the distributions of recovered $\nu_s$ obtained from the individual simulated spectra to test whether the finite scintle error (FSE) prescription introduced in Section~\ref{sec:FRB_pipeline} provides a reliable estimate of their uncertainty. For each $N_{\rm scint}$ case, we compare the measured scatter in $\nu_s$, estimated following the procedure described in Section~\ref{sec:uncertainties}, with the median of the fitting uncertainties and the FSE calculated using the known spectral bandwidth, $B$, and the input scintillation bandwidth, $\nu_{s,\rm input}$. 
To assess the dependence of the FSE estimate on the assumed filling factor, we consider three values, $\eta=0.1$, $0.2$, and $0.3$, spanning the commonly adopted range
\citep{cordes2010measurementmodelprecisionpulsar}. Fig~\ref{fig:fse_validation} shows how the agreement between the FSE and the measured scatter depends on the adopted value of $\eta$, with the discrepancy being most pronounced at low $N_{\rm scint}$ and decreasing as $N_{\rm scint}$ increases. At the same time, the FSE estimates are generally larger than the median formal uncertainties from the individual ACF fits, indicating that the fitting uncertainties alone are even more likely to underestimate the true measurement uncertainty. 
Given this difference, we use the measured scatter to empirically constrain the filling factor $\eta$. The resulting distributions are summarized in Fig~\ref{fig:eta_recovery} by their median and 15th--90th percentile intervals.

The recovered $\eta$ becomes increasingly stable with increasing $N_{\rm scint}$ and converges towards $\eta\simeq0.05$--$0.06$ for the better-constrained cases, below the commonly adopted value of $\eta=0.2$. For $N_{\rm scint}\leq25$, the measured scatter can exceed the limit $\sigma_{\nu_s}>\nu_{s,\rm input}/(2\ln2)$, resulting in negative inferred values of the filling factor. Since $\eta<0$ has no physical interpretation, this indicates that the FSE prescription cannot reproduce the simulated scatter in this regime.
Overall, the simulations indicate that the FSE becomes more representative as the number of scintles increases. This is because the standard FSE equation assumes Gaussian statistics, while at low $N_{\rm scint}$ the statistics are no longer Gaussian and the equation begins to fail. 

Based on the convergence obtained for the better-constrained cases, we conclude that $\eta=0.055$ could provide a reasonable empirical filling factor for the FSE estimates.   

\subsubsection{Uncertainty estimation at low $N_{\rm scint}$}
\label{sec:gamma_model}
\begin{figure}[H]
\centering

\begin{minipage}{0.45\textwidth}
\centering
\includegraphics[width=0.85\linewidth]{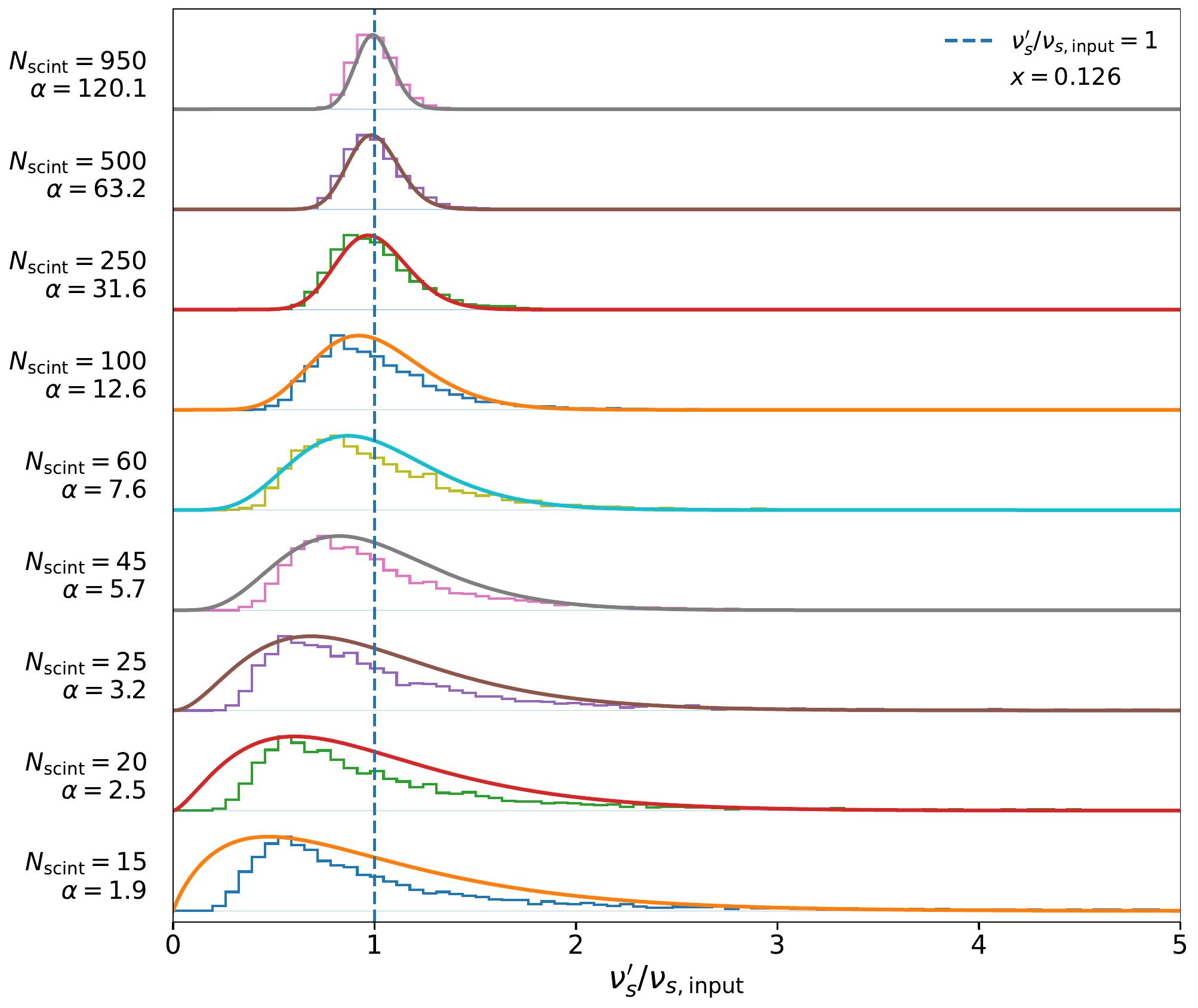}
\caption{Distributions of the recovered-to-input scintillation bandwidth ratio, $\nu_s'/\nu_{s,\rm input}$, for different simulated values of $N_{\rm scint}$. Each histogram shows the distribution obtained from the individual simulated realizations, while the solid curves show the corresponding Gamma model from the simultaneous fit to all $N_{\rm scint}$ cases. The shape parameter is given by $\alpha=xN_{\rm scint}$, with the best-fitting global parameter $x=0.126$. The vertical dashed line marks $\nu_s'/\nu_{s,\rm input}=1$, corresponding to perfect recovery of the input scintillation bandwidth. The distributions become progressively broader and more asymmetric toward low $N_{\rm scint}$, illustrating the increasing scatter in the recovered $\nu_s$.}
\label{fig:gamma_func}
\end{minipage}
\end{figure}

We explore a possible approach to estimate the uncertainty on $\nu_s$ at low $N_{\rm scint}$. For each simulated realization, we define the ratio between the recovered and input scintillation bandwidth as $r=\nu_s'/\nu_{s,\rm input}$, where $r=1$ corresponds to a perfect recovery of $\nu_{s,\rm input}$. At low $N_{\rm scint}$, we expect that the recovered values show a broad distribution around $r=1$, therefore a larger uncertainty on $\nu_s$, which becomes progressively narrower as $N_{\rm scint}$ increases. We describe these distributions using a Gamma distribution: 
\begin{equation}
f(r;\alpha,\theta)
=
\frac{1}{\Gamma(\alpha)\theta^\alpha}
r^{\alpha-1}
\exp\left(-\frac{r}{\theta}\right),
\label{eq:gamma_model}
\end{equation}
where $\alpha$ and $\theta$ are the shape and scale parameters, respectively. We set $\theta=1/\alpha$, which fixes the mean of the distribution to $\langle r\rangle=1$, and assume that the shape parameter increases linearly with the number of scintles, such that $\alpha=xN_{\rm scint}$. Here $x$ is a proportionality factor determined from the simulations. All simulated distributions are fitted simultaneously, with $x$ as a global free parameter. The normalised distributions and corresponding Gamma fits are shown in Fig~\ref{fig:gamma_func}, where we also report the best-fitting value of $x$.  
For the simulation setup considered here, the simultaneous fit gives the global parameter $x=0.126$. We use the Gamma model calibrated on the simulations as a possible prescription to estimate the uncertainties of the individual bursts in the observational data, after first evaluating the performance of the standard FSE in the corresponding $N_{\rm scint}$ regime. The calibrated Gamma distribution could then be used to construct, for each measured $\nu_s'$, a posterior distribution for the underlying true scintillation bandwidth $\nu_s$. The median of this distribution would provide an estimate of $\nu_s$, while the 16th and 84th percentiles would define its 68\% reliable interval.

\subsubsection{Expected number of scintles in FRB observations}
\label{sec:real_obs_scintles}
To assess whether the number of scintles explored in our simulations is representative of real FRB observations, we consider three lines of sight away from the Galactic plane and use NE2025 to estimate the Milky Way contribution to the scintillation bandwidth up to a distance of $1~\mathrm{kpc}$ along each line of sight. We selected these off-plane directions by looking at the sky distribution of CHIME FRBs in \citet{Patil_2026}. We obtain $\nu_s \sim 5$--$10$~MHz at 1~GHz and scale these values, assuming the Kolmogorov relation $\nu_s\propto\nu^{4.4}$, to the observing bands of three representative FRB instruments: LOFAR (110--188~MHz) \citep{Chawla_2025}, CHIME (400--800~MHz) \citep{Patil_2026}, and DSA (880--980, 1310--1510, and 1650--1950~MHz) \citep{connor2026dsachronoscopefastradioburst}. At LOFAR and CHIME frequencies, $\nu_s$ is much smaller than the observing bandwidth, corresponding to $N_{\rm scint}$ of order $\sim10^4$ and $\sim10^2$--$10^3$, respectively. At DSA frequencies, however, the rapid increase of $\nu_s$ progressively reduces $N_{\rm scint}$. In this regime, $\nu_s$ becomes comparable to the available bandwidth and the ACF is determined by a limited number of scintles. This highlights the importance of the ratio $B/\nu_s$ when interpreting the ACF and estimating the uncertainty on $\nu_s$.
For FRBs closer to the Galactic plane stronger scattering is expected to produce narrower scintillation bandwidths hence a larger number of scintles is expected.

\subsubsection{Individual bursts of \rnew}
\label{sec:r147_singlebursts}
We now present the results from modelling the ACFs of individual bursts of \rnew\ observed with the Effelsberg UBB receiver in a specific epoch of hyper-activity. 
Here, we focus only on the analysis performed on the bursts observed in Band~2, covering approximately 1940--2589~MHz. In particular, we select the 33 bursts with the highest S/N in this band in order to reduce the influence of noise on the ACF measurements.

For each burst, we manually select the frequency range containing the emission and remove channels affected by RFI. This provides a burst-specific usable bandwidth, determined by the spectral extent of the emission and the excluded channels. 
We then compute the frequency ACF using Equation~\ref{eq:acf} and apply the fitting procedure described in Section~\ref{sec:FRB_pipeline} to measure the scintillation bandwidth.

\begin{figure*}[ht]
\centering

\includegraphics[width=0.9\textwidth]{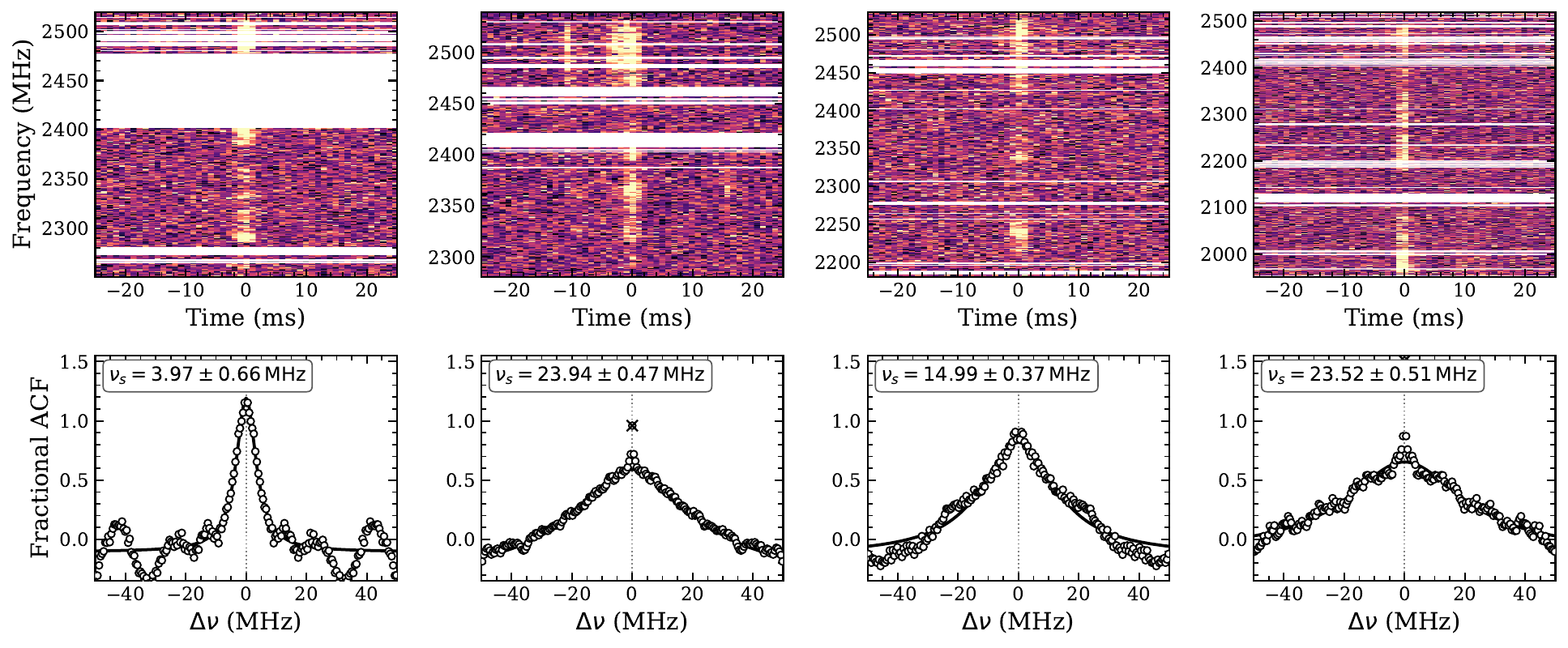}
\caption{Example bursts from \rnew\ observed in UBB Band~2 and their corresponding frequency ACFs. The black curves show the best-fitting Lorentzian models. The scintillation bandwidth, $\nu_s$, obtained from each individual ACF and its associated formal fit uncertainty are reported in each panel.}
\label{fig:observed_acf}
\end{figure*}

In Fig~\ref{fig:observed_acf}, we present example dynamic spectra and their corresponding ACFs for \rnew. 
They illustrate the diversity of band-limited spectral morphologies and the corresponding variations in the shape and quality of the measured ACFs. Bursts with narrower usable bandwidths and noisier spectra generally produce less smooth ACFs, with prominent fluctuations that can make the Lorentzian fit less reliable. 
In contrast, bursts spanning a larger usable bandwidth tend to produce smoother ACFs, consistent with the finite-bandwidth effects found in our simulations. However, this behaviour is less straightforward in real data, where additional effects such as noise, RFI, and intrinsic spectral structure can complicate the ACF and its interpretation even for bursts with broad frequency coverage. 

In Fig~\ref{fig:observed_acf}, we report only the formal uncertainty associated with the fitting procedure for each individual measurement of $\nu_s$. We do not apply the FSE at this stage. In the following section, we investigate whether combining the available measurements provides a more stable estimate of the characteristic $\nu_s$, which can then be used to assess the applicability of the FSE prescription to the observations.

\subsection{ACF averaging across multiple bursts}
\label{sec:averaging_results}
We now consider the case where multiple bursts are available and investigate how combining their individual ACFs can improve the measurement of the scintillation bandwidth.
\subsubsection{Simulations}
\label{sec:averaging_simulations}
The bottom panels of Fig.~\ref{fig:simulations} illustrate the effect of averaging 50 independent ACFs for two very different regimes, with $N_{\rm scint}=15$, representative of our observations, and $N_{\rm scint}=950$, representing the ideal case of dense scintillation structures across the observing bandwidth. As discussed above, 50 realizations were chosen to represent a realistic number of bursts that may be available in observations. Averaging reduces the fluctuations of the individual ACFs and produces a smoother profile that can be more reliably modelled.
The improvement is particularly evident for $N_{\rm scint}=15$, where individual ACFs show substantial fluctuations and the characteristic wiggles. Despite this challenging regime, the averaged ACF approaches a Lorentzian profile and is well modelled by the fitting procedure. Instead, for $N_{\rm scint}=950$, the individual ACFs are already considerably smoother, and their average closely follows the expected profile.

We note that bursts occurring within the same scintillation timescale are expected to show correlated ACF structures. However, our simulations focus on independent realizations to investigate how averaging uncorrelated ACFs reduces fluctuations. These results demonstrate that averaging multiple independent ACFs can mitigate finite-scintle variability and improve the recovery of $\nu_s$, supporting its application to repeating FRBs for which multiple burst spectra are available.

\subsubsection{Analysis of \rnew}
\label{sec:averaging_r147}
Motivated by the improved stability obtained by averaging the simulated ACFs, we now apply the same approach to the observational data of \rnew. Our dataset consists of six hours of observations, spanning a few predicted scintillation timescales. While the scintillation of some close-by bursts can be correlated, the whole sample of ACFs contains sufficiently independent data such that their average has reduced fluctuations and provides a more stable estimate of the scintillation bandwidth. This time, to account for the fact that the observed bursts have different bandwidths, we obtain a weighted average of their ACFs, defined as:

\begin{equation}
\overline{C}(\Delta\nu)
=
\frac{
\sum_i W_i(\Delta\nu)\,C_i(\Delta\nu)}{\sum_i W_i(\Delta\nu)},
\label{eq:weighted_mean}
\end{equation}  

\begin{equation}
W_i(\Delta\nu) = N_{{\rm pair},i}(\Delta\nu)\,
\label{eq:weight}
\end{equation}
with $N_{{\rm pair},i}(\Delta\nu)$ the number of valid channel pairs contributing to the ACF of burst $i$ at frequency lag $\Delta\nu$. This technique gives greater weight to ACF measurements supported by a larger usable spectral range, where the single modelling is supposed to be more reliable. We note that, since the weights depend on $\Delta\nu$, the relative contribution of individual bursts varies across the ACF, which may introduce small changes in the shape of the averaged ACF. 

The weighted average ACF is presented in the bottom panel of Fig~\ref{fig:bootstrap}. We then apply the fitting procedure described in Section~\ref{sec:FRB_pipeline} to the weighted average ACF and determine the scintillation bandwidth from the best-fitting Lorentzian model, denoting the value obtained from the weighted-average ACF as $\overline{\nu}_s$. Despite the fluctuations introduced by the individual ACFs, their combination produces a more stable central ACF structure and therefore a more robust estimate of the characteristic scintillation bandwidth. Importantly, averaging bursts obtained within a single observing session is not expected to bias the inferred scintillation bandwidth, since significant changes in the scintillation pattern are expected on the longer refractive timescale, when the effective line of sight samples a different region of the scattering screen.

Modelling the weighted average ACF in UBB Band~2 with the single-Lorentzian model of Equation~\ref{eq:lorentzian} yields a scintillation bandwidth of $\overline{\nu}_s \simeq 17.50$~MHz, as reported in the bottom panel of Fig~\ref{fig:bootstrap}. The NE2025 prediction of $\nu_s \simeq 0.52$~MHz at 1~GHz corresponds to $\simeq 18.96$~MHz at the central frequency of Band~2 ($2.2645$~GHz), assuming the standard Kolmogorov scaling $\nu_s \propto \nu^{4.4}$. Our measurement is therefore in good agreement with the NE2025 prediction in this band. 

However, we note that this weighted-average ACF shows a narrow central peak that is not fully captured by the single-Lorentzian model. We therefore also explored a double-Lorentzian description of the form 
\begin{equation}
{\rm ACF}(\Delta\nu)=
\frac{A_1}{1+(\Delta\nu/\nu_{s,1})^2}+\frac{A_2}{1+(\Delta\nu/\nu_{s,2})^2}+C, 
\label{eq:double_lorentzian}
\end{equation}
here $\nu_{s,1}$ and $\nu_{s,2}$ represent two characteristic scintillation scales. A more complete two-screen treatment can also include a mixed term arising from the combined modulation of the two screens \citep{Pradeep2025}, which is not included in our two-Lorentzian model. 

Hence, the presence of two distinct scales may indicate scintillation arising from more than one scattering screen. In such a scenario, one scale is usually associated with scintillation in the Milky Way, while the other could originate from a screen closer to the FRB or within its host environment. 
Here, we demonstrate that a two-screen scenario is favoured by the ACF analysis. We measure scintillation scales of $\overline{\nu}_{s,1} = 2.11$~MHz and $\overline{\nu}_{s,2} = 20.60$~MHz. The broader scale remains consistent with the NE2025 prediction, while the origin of the narrower scale requires further investigation. 

A comparison of the single- and double-Lorentzian models is shown in the bottom panel of Fig.~\ref{fig:bootstrap}. Detailed modelling is left for future work; here, we primarily focus on demonstrating that, despite the variability observed among individual bursts, the weighted-average ACF remains stable. Determining the most appropriate model will ultimately depend on the scintillation scenario that best describes the observed ACF behaviour.

\subsection{Uncertainty of the observational measurements}
\label{sec:uncertainties_observational_data}
\subsubsection{Individual burst uncertainties}
\label{sec:error_bursts_observational}
In the previous subsection, we discussed that the weighted average ACF provides a more stable estimate of the characteristic scintillation bandwidth. We therefore use its value together with the usable bandwidth of each burst, $B_i$, to estimate their corresponding number of scintles as $N_{{\rm scint,mean},i} = \frac{B_i}{\overline{\nu}_s}$. We find that most bursts fall in the low-$N_{\rm scint}$ regime, where our simulations show that the FSE does not reliably describe the uncertainty on $\nu_s$.

Consequently, rather than applying the FSE in this regime, we use the Gamma distribution calibrated on the simulations, as reported in Subsection~\ref{sec:gamma_model}. Before applying the Gamma model, we test whether the scatter predicted by the simulation is consistent with that observed in the data. We restrict this comparison to the 22 bursts with reliable individual ACF fits, excluding the cases around $\nu_s'\sim50$~MHz for which the fitting procedure does not reliably constrain the noisier ACFs.  From the variance of the Gamma distribution, the expected uncertainty for each burst is
\begin{equation}
\sigma_{\Gamma,i} = \frac{\overline{\nu}_s}{\sqrt{xN_{{\rm scint},i}}}
\label{eq:sigma_gamma}
\end{equation}
where $N_{{\rm scint},i}=B_i/\overline{\nu}_s$, $\overline{\nu}_s=17.49$~MHz is obtained from the weighted-average ACF and $x=0.126$ is derived from the simulations. We then calculate: 
\begin{equation}
\chi^2
= \sum_i \left(\frac{\nu_{s,i}'-\overline{\nu}_s} {\sigma_{\Gamma,i}}\right)^2.
\label{eq:chi2}
\end{equation}
For the $N = 22$ bursts retained in this comparison, we obtain $\chi^2\simeq12.01$, or $\chi^2/N\simeq0.55$. We therefore consider the Gamma-derived uncertainties to provide an acceptable description of the observed scatter, although the value below unity indicates that the predicted scatter is larger than that observed in the data.
For a given burst, the measured scintillation bandwidth $\nu_s'$ and the usable bandwidth $B$ are fixed. We then consider a range of possible values for the true $\nu_s$ and calculate the corresponding $N_{\rm scint}$ and the Gamma shape parameter $\alpha = x N_{\rm scint}$, considering $x=0.126$ as obtained from the simulations (see also Subsection~\ref{sec:gamma_model} and Fig~\ref{fig:gamma_func}).
The likelihood of measuring $\nu_s'$ for a given true $\nu_s$ is then

\begin{equation}
p(\nu_s' \mid \nu_s,B)
= \frac{1}{\nu_s} \frac{\alpha^\alpha}{\Gamma(\alpha)} \left(\frac{\nu_s'}{\nu_s}\right)^{\alpha-1} \exp\left(-\alpha\frac{\nu_s'}{\nu_s}\right).
\label{eq:likelihood}
\end{equation}

Assuming a flat prior, such that all allowed values of $\nu_s$ are initially considered equally likely, we normalise this likelihood to obtain the posterior:
\begin{equation}
p(\nu_s \mid \nu_s',B) = \frac{p(\nu_s' \mid \nu_s,B)}{\int p(\nu_s' \mid \nu_s,B)\,d\nu_s}.
\label{eq:posterior}
\end{equation} 

\begin{figure}[H]
\centering

\begin{minipage}{0.47\textwidth}
\centering
\includegraphics[width=0.83\linewidth]{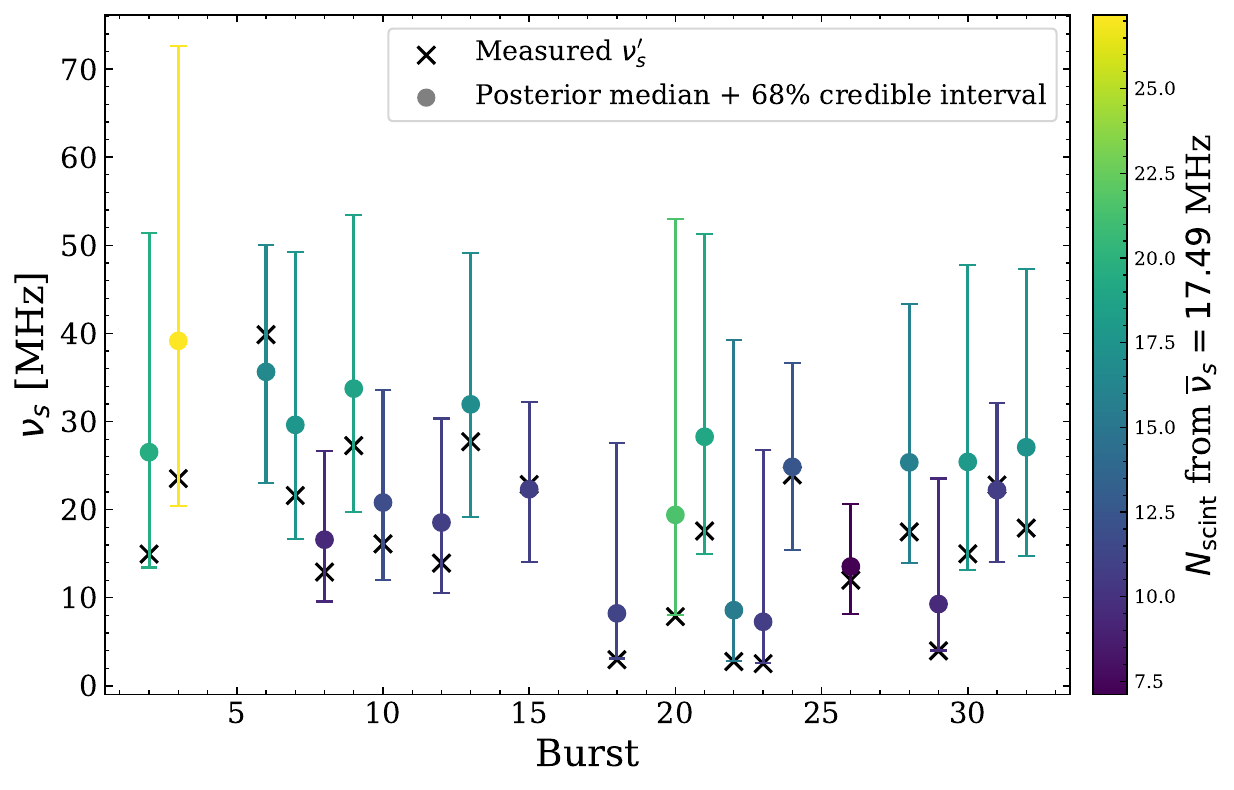}
\caption{Individual scintillation bandwidth measurements and posterior estimates for the bursts in UBB Band~2. Black crosses show the measured scintillation bandwidths $\nu_s'$, with marker size scaling with burst S/N. The coloured points show the median of the posterior distribution for the underlying $\nu_s$, with the error bars indicating the corresponding 68\% reliable intervals (16th--84th percentiles). The colour indicates $N_{\rm scint}$, calculated using the usable bandwidth of each burst and $\overline{\nu}_s=17.49$~MHz obtained from the weighted-average ACF. The posterior estimates are generally consistent with the measured values, with broad and asymmetric intervals reflecting the larger uncertainty expected in the low-$N_{\rm scint}$ regime.}
\label{fig:snr_nus}
\end{minipage}
\end{figure}

We take the posterior median as the inferred $\nu_s$, with the 16th and 84th percentiles defining its 68\% reliable interval. The resulting estimates are shown in Fig~\ref{fig:snr_nus}, together with the measured $\nu_s'$ values. The posterior median values are often shifted towards larger $\nu_s$, reflecting the asymmetric posterior distributions resulting from the Gamma-based inference. For a very few bursts, the measured $\nu_s'$ lies above the posterior median. This may be influenced by the adopted condition $N_{\rm scint}\geq5$, which, through $N_{\rm scint}=B/\nu_s$, imposes the burst-dependent upper limit  $\nu_s\leq B/5$ and restricts the high-$\nu_s$ side of the posterior. Overall, the posterior uncertainties are broad and asymmetric, reflecting the scatter found in the simulations. 

An alternative way to estimate these uncertainties would be to simulate the specific bandwidth, RFI, and noise properties of each burst, which we plan to explore in future work.

\subsubsection{Uncertainty on the average ACF}
\label{sec:bootstrapping_error}
\begin{figure}[H]
\centering

\begin{minipage}[t]{0.35\textwidth}
\centering
\includegraphics[width=\textwidth]{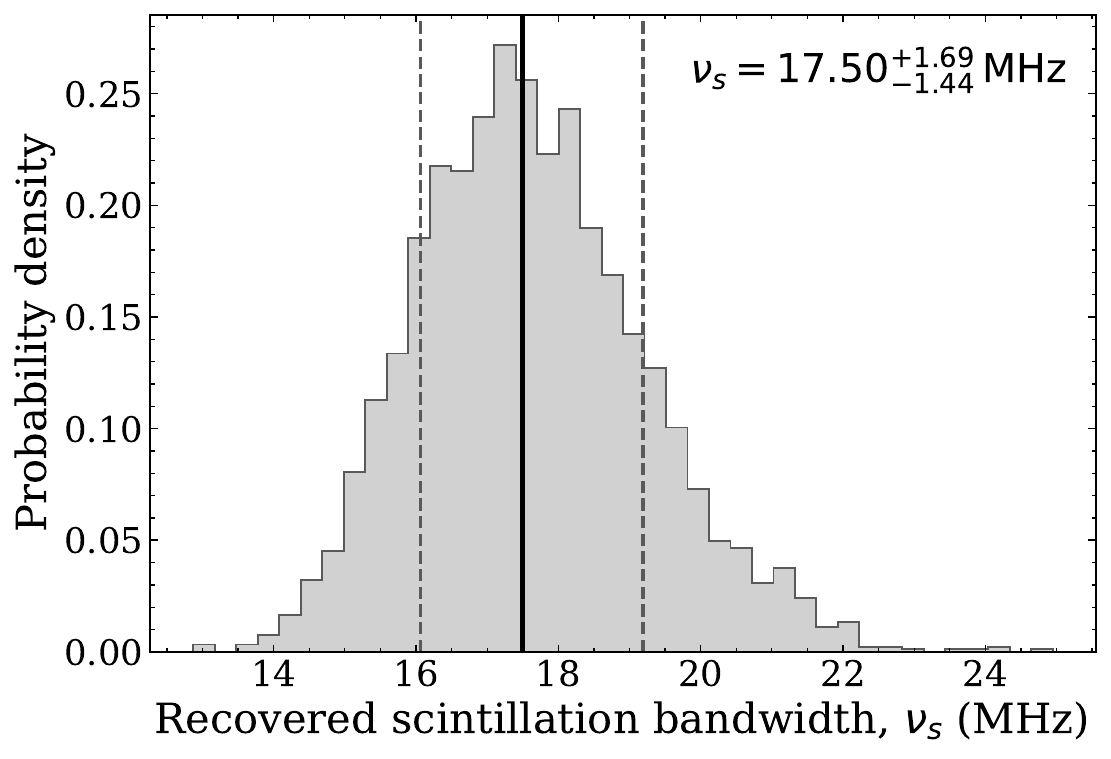}
\end{minipage}
\hfill
\begin{minipage}[t]{0.45\textwidth}
\centering
\includegraphics[width=\textwidth]{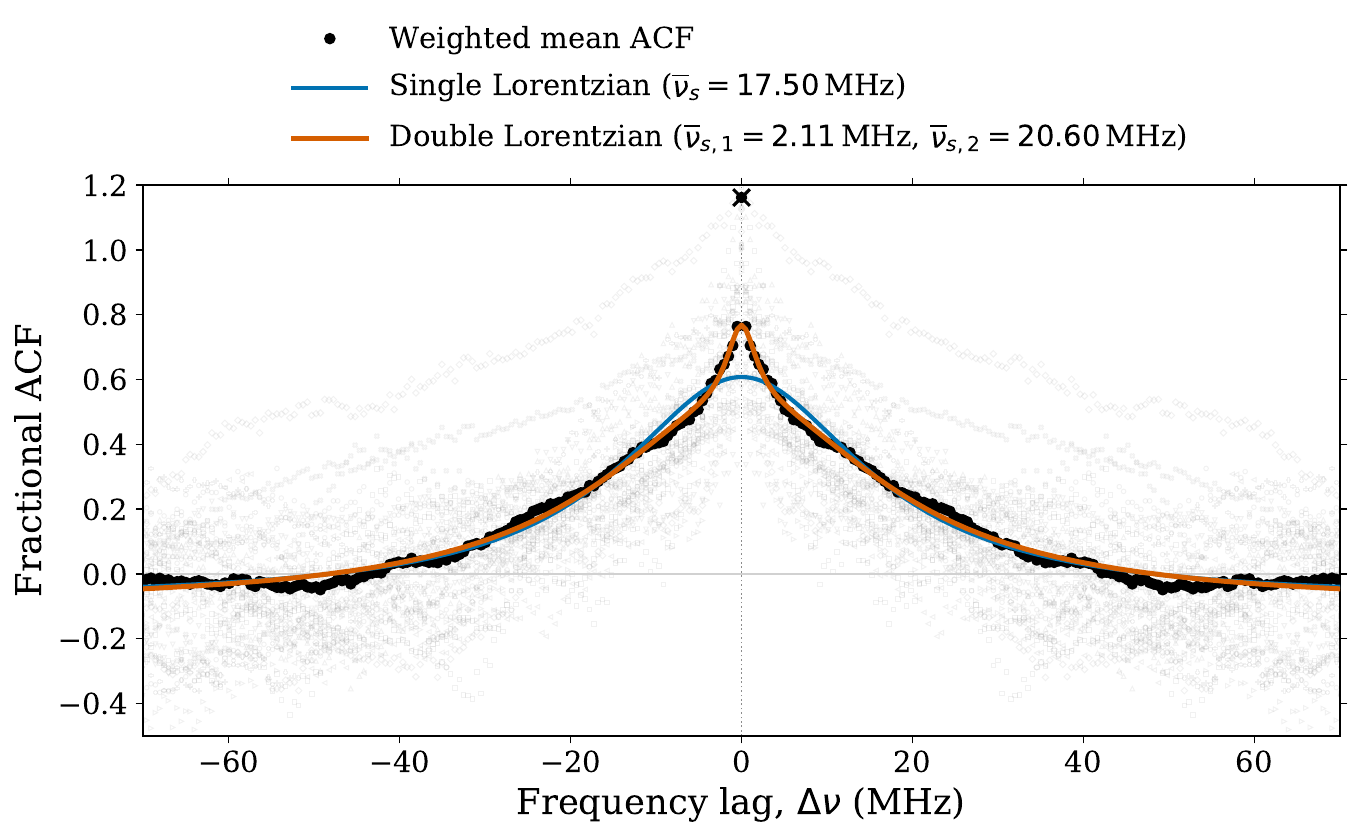}
\end{minipage}

\caption{Top:Bootstrap distribution of the scintillation bandwidth recovered from the weighted-average ACF using the single-Lorentzian model. The dashed lines indicate the 16th and 84th percentiles of the recovered $\overline{\nu}_s$ distribution. Bottom: Weighted-average ACF obtained from 33 bursts in UBB Band~2, shown in black, with the ACFs of the individual bursts displayed in the background. The best-fitting single- and double-Lorentzian models are overlaid for comparison. The single-Lorentzian fit provides a single characteristic scintillation bandwidth, $\overline{\nu}_s$, whereas the double-Lorentzian model yields two characteristic scintillation scales, $\overline{\nu}_{s,1}$ and $\overline{\nu}_{s,2}$.
}
\label{fig:bootstrap}

\end{figure}

For the weighted average ACF, the uncertainty is estimated separately. Since the contributing bursts cover different frequency ranges, a single representative bandwidth $B$ cannot be unambiguously defined for the combined measurement, preventing a direct application of the FSE prescription. We therefore estimate the uncertainty empirically from the burst sample using bootstrap resampling. In each bootstrap iteration, we resample the same number of bursts as in the original sample, construct the weighted average ACF, and repeat the Lorentzian fit. We perform 3000 iterations and use the 16th and 84th percentiles of the resulting $\overline{\nu}_s$ distribution to define the uncertainty on the combined measurement, reflecting the variation in $\overline{\nu}_s$ across the different bootstrap realizations. The resulting distribution is shown in the top panel of Fig~\ref{fig:bootstrap}.

\subsection{Effect of the number of averaged ACFs}
\label{sec:number_of_spectra}
Additionally, we use the simulations to test how the number of spectra included in the average affects the recovered scintillation bandwidth. For each $N_{\rm scint}$ case, we average different numbers of ACFs, $N_{\rm group}$, and fit the resulting mean ACF with the same Lorentzian model used for the individual spectra. We then compare the recovered $\overline{\nu}_s$ with the known input value, $\nu_{s,\rm input}$.

\begin{figure*}
\centering
\includegraphics[width=13.5cm]{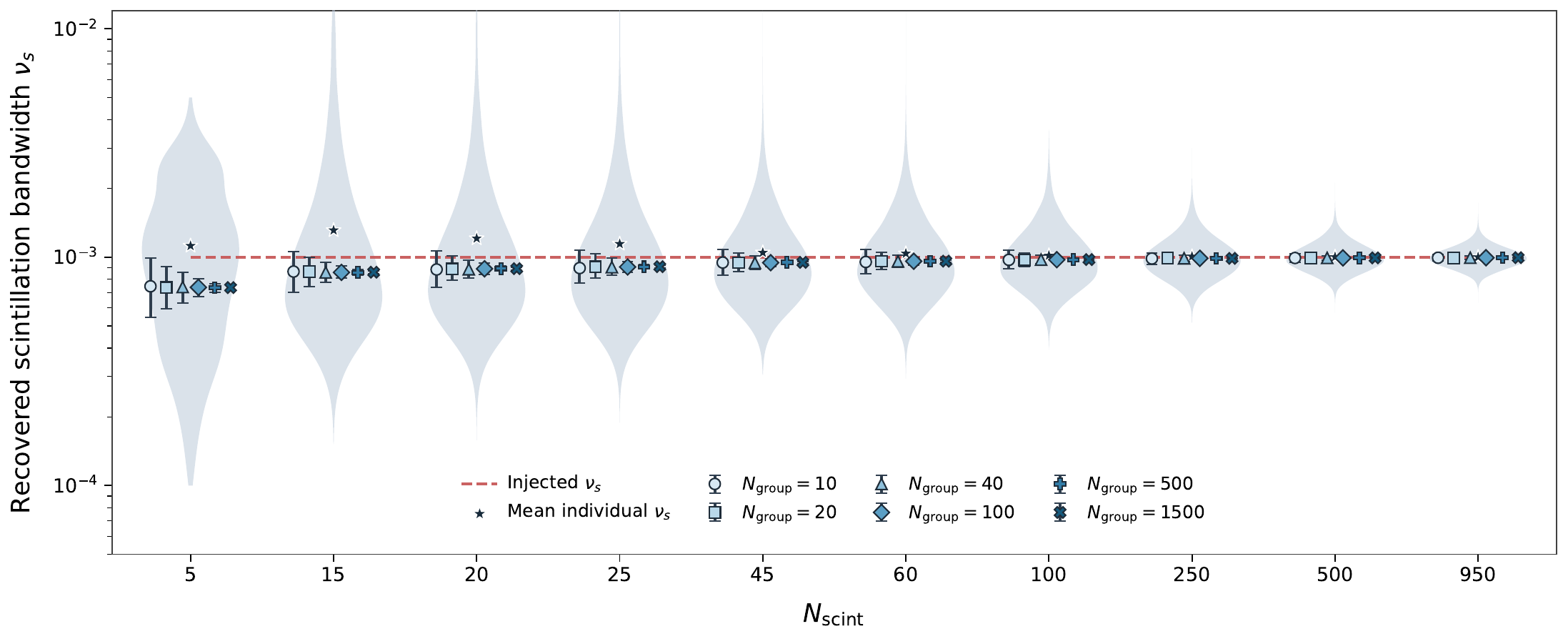}
\caption{Recovered scintillation bandwidth,$\nu_s$, as a function of $N_{\rm scint}$. Each violin represents the distribution obtained from the 20000 individual simulated realizations, with its width indicating the relative concentration of recovered values. The star indicates the mean $\nu_s$ recovered from the individual realizations. The different symbols indicate the $\overline{\nu}_s$ values recovered from mean ACFs constructed using different group sizes, $N_{\rm group}$. The dashed horizontal line indicates the injected $\nu_{\rm s,input}$.} 
     \label{fig:violin}
\end{figure*}

Fig~\ref{fig:violin} shows the $\nu_s$ values recovered from the individual spectra and from the averaged ACFs for different $N_{\rm group}$. As more spectra are averaged, the recovered $\overline{\nu}_s$ becomes more stable and converges towards the input value. This improvement benefits the low $N_{\rm scint}$ cases, where the individual measurements exhibit a larger scatter. This test is relevant for our observations, where the number of bursts available for averaging is limited by the S/N, usable bandwidth, and RFI contamination. In most UBB bands, we used approximately 33 selected bursts. Although this is a relatively small sample, our simulations show that averaging a similar number of spectra can already provide a stable estimate of $\nu_s$. A larger sample of bursts, particularly with high S/N and especially broad usable bandwidth, would further improve the measurement.

\subsection{Modulation index of simulated spectra}
\label{sec:mod_index}
\begin{figure}[H]
\begin{minipage}{0.49\textwidth}
\centering
\includegraphics[width=0.83\textwidth]{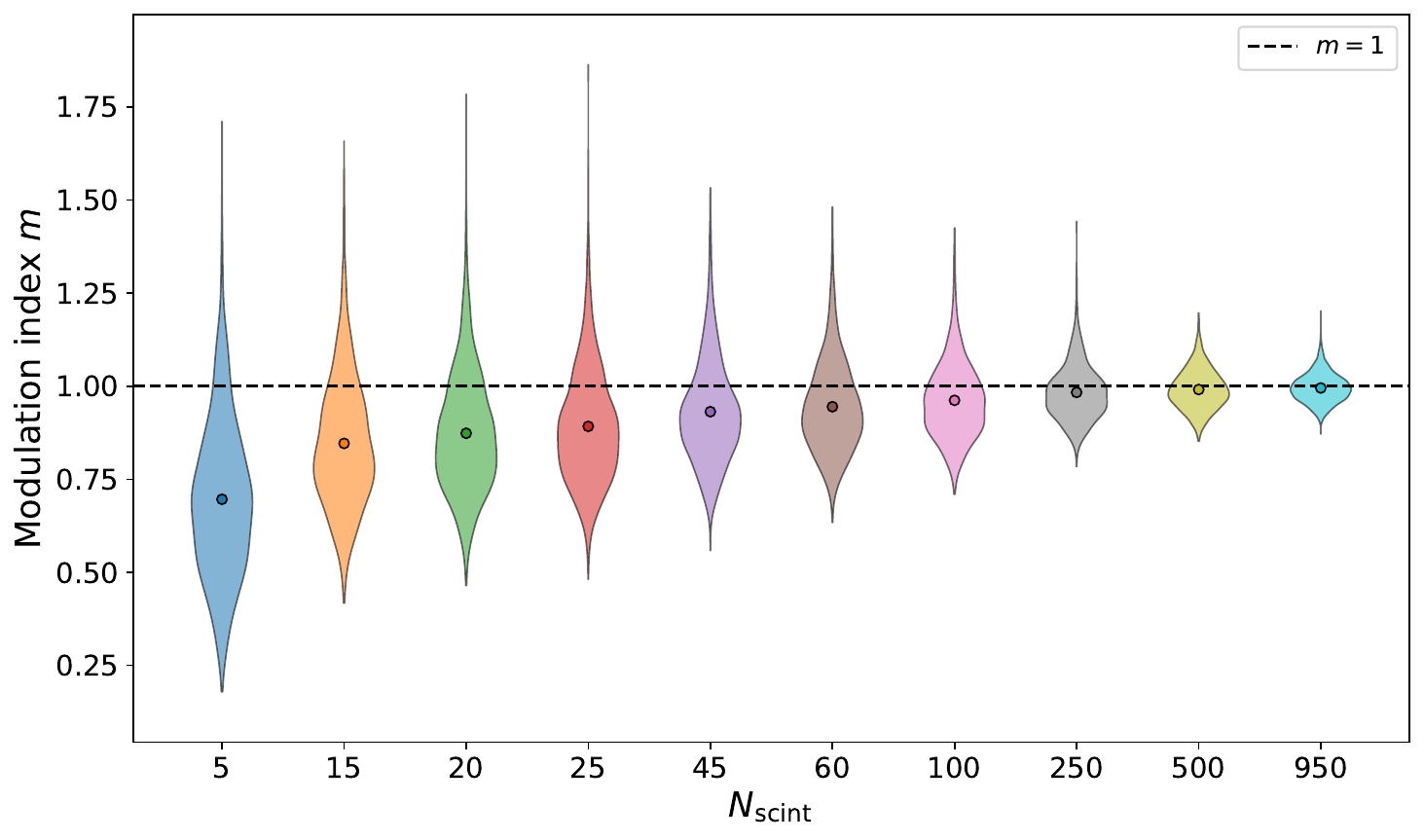}

\caption{Modulation index $m$ as a function of the number of scintles, $N_{\rm scint}$, measured from the 1500 individual simulated realizations. The violin plots show the distribution of the modulation index across the realizations for each $N_{\rm scint}$, while the black-edged markers indicate the mean value of each distribution. The dashed horizontal line marks the expected value $m=1$ for fully modulated scintillation. The distributions become progressively narrower with increasing $N_{\rm scint}$, with the mean modulation index approaching unity.}\label{fig:mod_index}
\end{minipage}
\end{figure}

Finally, as a further diagnostic, we investigate the behaviour of the recovered modulation index, $m$, as a function of $N_{\rm scint}$ in the simulated spectra. 
For each value of $N_{\rm scint}$, we use 1500 independent realizations and estimate the modulation index from the fitting procedure according to Equation~\ref{eq:modulation_index}. We then calculate the mean modulation index across the realizations for each $N_{\rm scint}$. Since our simulations describe scintillation produced by a single scattering screen, the modulation index is expected to approach $m \simeq 1$. 
Fig~\ref{fig:mod_index} shows that the number of scintles contained within the available bandwidth can affect the recovered modulation index, particularly at low $N_{\rm scint}$, where both a larger scatter and a deviation of the mean from unity are observed. As $N_{\rm scint}$ increases, the distribution becomes progressively narrower and the mean approaches $m \simeq 1$. We include this result as a consistency check and to highlight the potential effect of a limited number of scintles within the observing bandwidth on modulation-index measurements. 

\section{Conclusions}
\label{sec:conclusion}
In this work, we investigated the challenges and uncertainties associated with modelling the scintillation frequency scale in FRBs when the available observing bandwidth is limited. Using controlled simulations with a known input scintillation bandwidth, we tested both individual ACF measurements and the averaging of multiple independent ACFs. We then applied the same analysis to observations of the repeating \rnew\ obtained with the Effelsberg UBB receiver. Our main conclusions are:

\begin{itemize}
\renewcommand{\labelitemi}{$\bullet$}
\vspace{-0.3cm}
\item The reliability of individual ACFs measurements strongly depends on the number of scintillation features, or scintles, across the bandwidth. As $N_{\rm scint}$ decreases, the ACF develops more prominent wiggles and deviates from the expected Lorentzian profile, leading to a larger scatter in the recovered $\nu_s$. Therefore, such ACF structure can arise naturally from isolated scintle cases and does not necessarily imply additional complexity in the scattering medium.

\item From the simulations we found that the FSE prescription generally underestimates the scatter recovered from the individual simulated spectra when the commonly adopted filling factor $\eta\sim0.1$--$0.3$ is used. In the better-constrained, higher-$N_{\rm scint}$ cases, the empirically recovered filling factor converges towards $\eta\simeq0.05$--$0.06$. At $N_{\rm scint}\leq 25$, the scatter of the simulated ACFs is too large to be reproduced by any physical value of $\eta$, indicating a limitation of the FSE prescription in the low-scintle regime.

\item In the low-$N_{\rm scint}$ regime, we used the simulations to calibrate a Gamma model for the recovered $\nu_s$ and we demonstrate that it provides reasonable estimates of the measurement uncertainties. We then applied this model to the observational data to infer the posterior distribution of the true $\nu_s$ with the uncertainties determined from its 16th and 84th percentiles, corresponding to the 68\% credible interval.

\item When multiple independent spectra from a single source are available, averaging their ACFs improves the stability of the recovered scintillation bandwidth. This improvement is particularly important at low $N_{\rm scint}$, where individual measurements show the largest scatter. As more ACFs are included in the average, the recovered scintillation bandwidth progressively approaches the known input value, demonstrating the advantage of this approach for repeating FRBs.

\item Applying the averaging approach to the Effelsberg UBB observations of \rnew\ produces a more stable ACF than those obtained from individual bursts. In UBB Band~2, the weighted average ACF yields $\overline{\nu}_s\simeq17.50$~MHz at 2.2645~GHz, in good agreement with the NE2025 prediction at the same frequency. A two-Lorentzian model may provide a more suitable description; however, the exact modelling is left for future work. Here, we limit our analysis to demonstrating the stability of the weighted-average ACF as a basis for further investigation.

\item We further find that the recovered modulation index approaches the expected value of $m \simeq 1$ as $N_{\rm scint}$ increases, while a limited number of scintles leads to larger scatter and values that are systematically lower than unity. 
\end{itemize}
\vspace{-0.3cm}
Overall, our results demonstrate that the number of scintles within the available bandwidth is a key factor in the reliable measurement of scintillation frequency scales in FRBs. For individual bursts, the FSE could provide a practical uncertainty estimate when a sufficient number of scintles is present, with our simulations favouring $\eta\simeq0.05$--$0.06$. At low $N_{\rm scint}$, where the FSE becomes unreliable, 
we instead use a simulation-calibrated Gamma prescription to account for the scatter in the recovered $\nu_s$. The resulting credible intervals are broader than the formal least-squares uncertainties and provide a more realistic estimate of the measurement uncertainty. We note that the uncertainties discussed here are systematic and do not include the contribution from thermal noise, which would introduce an additional uncertainty.
For repeating FRBs, averaging multiple independent ACFs improves the recovery of $\overline{\nu}_s$, while bootstrap resampling provides a natural way to estimate the uncertainty when bursts span different bandwidths. Our results suggest that this framework could be useful for future FRB observational studies, where its application will help assess its effectiveness across different observing conditions and scintillation regimes.

\begin{acknowledgements}
This work is based on observations with the 100-m telescope of the MPIfR (Max-Planck-Institut für Radioastronomie) at Effelsberg. The authors thank Dr. Alex Kraus for scheduling observations with the new UBB receiver. The Effelsberg UBB receiver and EDD backend system are developed and maintained by the Electronics Division at the MPIfR.

The authors thank Robert A. Main and Sachin Pradeep E.T. for insightful discussions and valuable input. LN acknowledges Robert Reischke for providing useful tools, and Adarsh Bathula for constructive feedback on the scintillation modelling approach, additional tests, and complementary analyses. LN also thanks R. Shannon, C. James, and A. Deller for their suggestions and interest in this work as well as her colleagues in the Lise Meitner Research Group for stimulating discussions and comments. We sincerely thank the referee for the positive feedback and valuable suggestions that helped improve the manuscript. LGS is a Lise Meitner Research Group Leader and acknowledges funding from the Max Planck Society.
\end{acknowledgements}

\vspace{-0.6cm}
\bibliography{references}

@ARTICLE{Lorimer2007,
       author = {{Lorimer}, D.~R. and {Bailes}, M. and {McLaughlin}, M.~A. and {Narkevic}, D.~J. and {Crawford}, F.},
        title = "{A Bright Millisecond Radio Burst of Extragalactic Origin}",
      journal = {Science},
         year = 2007,
        month = nov,
       volume = {318},
       number = {5851},
        pages = {777},
          doi = {10.1126/science.1147532},
archivePrefix = {arXiv},
       eprint = {0709.4301},
 primaryClass = {astro-ph},
       adsurl = {https://ui.adsabs.harvard.edu/abs/2007Sci...318..777L}
}

@article{Abbott2026,
doi = {10.3847/1538-4365/ae3828},
url = {https://doi.org/10.3847/1538-4365/ae3828},
year = {2026},
month = {mar},
publisher = {The American Astronomical Society},
volume = {283},
number = {1},
pages = {34},
author = {The {CHIME/FRB Collaboration} and Abbott, Thomas and Andersen, Bridget C. and Andrew, Shion and Bandura, Kevin and Bhardwaj, Mohit and Bhusare, Yash and Brar, Charanjot and Cassanelli, Tomas and Chatterjee, Shami and Cliche, Jean-Francois and Cook, Amanda M. and Curtin, Alice and Dobbs, Matt and Dong, Fengqiu Adam and Eadie, Gwendolyn and Eftekhari, Tarraneh and Fonseca, Emmanuel and Gaensler, B. M. and Good, Deborah and Halpern, Mark and Hessels, Jason W. T. and Ibik, Adaeze and Jain, Naman and Joseph, Ronniy C. and Kader, Zarif and Kaspi, Victoria M. and Khan, Afrokk and Kharel, Bikash and Kumar, Ajay and Landecker, T. L. and Lang, Dustin and Lanman, Adam E. and L’Argent, Magnus and Lazda, Mattias and Leung, Calvin and Li, Dong Zi and Lintott, Chris J. and Main, Robert and Masui, Kiyoshi W. and Mate, Sujay and McGregor, Kyle and Mckinven, Ryan and Mena-Parra, Juan and Meyers, Bradley W. and Michilli, Daniele and Ng, Cherry and Ng, Mason and Nimmo, Kenzie and Noble, Gavin and Pandhi, Ayush and Patil, Swarali S. and Pearlman, Aaron B. and Pen, Ue-Li and Pleunis, Ziggy and Prochaska, J. Xavier and Rafiei-Ravandi, Masoud and Ransom, Scott and Renard, Andre and Sammons, Mawson W. and Sand, Ketan R. and Scholz, Paul and Shah, Vishwangi and Shin, Kaitlyn and Siegel, Seth R. and Sirota, Sloane and Smith, Kendrick and Stairs, Ingrid and Stenning, David C. and Tendulkar, Shriharsh P. and Vanderlinde, Keith and Walmsley, Mike and Wang, Haochen and Wulf, Dallas},
title = {The Second CHIME/FRB Catalog of Fast Radio Bursts},
journal = {\apjs}
}

@ARTICLE{Andersen2019,
       author = {{CHIME/FRB Collaboration} and {Andersen}, B.~C. and {Bandura}, K. and {Bhardwaj}, M. and {Boubel}, P. and {Boyce}, M.~M. and {Boyle}, P.~J. and {Brar}, C. and {Cassanelli}, T. and {Chawla}, P. and et al.},
        title = "{CHIME/FRB Discovery of Eight New Repeating Fast Radio Burst Sources}",
      journal = {\apjl},
         year = 2019,
        month = nov,
       volume = {885},
       number = {1},
          eid = {L24},
        pages = {L24},
          doi = {10.3847/2041-8213/ab4a80},
archivePrefix = {arXiv},
       eprint = {1908.03507},
 primaryClass = {astro-ph.HE},
       adsurl = {https://scixplorer.org/abs/2019ApJ...885L..24C}
}

@article{Bochenek2020,
   title={A fast radio burst associated with a Galactic magnetar},
   volume={587},
   ISSN={1476-4687},
   url={http://dx.doi.org/10.1038/s41586-020-2872-x},
   DOI={10.1038/s41586-020-2872-x},
   number={7832},
   journal={Nature},
   publisher={Springer Science and Business Media LLC},
   author={Bochenek, C. D. and Ravi, V. and Belov, K. V. and Hallinan, G. and Kocz, J. and Kulkarni, S. R. and McKenna, D. L.},
   year={2020},
   month=Nov, pages={59–62} }

@ARTICLE{lower2026transientnarrowbandradiobursts,
       author = {{Lower}, Marcus E. and {Scholz}, Paul and {Camilo}, Fernando and {Palmer}, David M. and {Reynolds}, John E. and {Sarkissian}, John M. and {Toomey}, Lawrence J. and {Younes}, George},
        title = "{Transient narrowband radio bursts from the magnetar 1E 1547.0-5408}",
      journal = {arXiv e-prints},
         year = 2026,
        month = mar,
          eid = {arXiv:2603.21450},
        pages = {arXiv:2603.21450},
          doi = {10.48550/arXiv.2603.21450},
archivePrefix = {arXiv},
       eprint = {2603.21450},
 primaryClass = {astro-ph.HE},
       adsurl = {https://scixplorer.org/abs/2026arXiv260321450L}
}

@ARTICLE{Bannister2019,
       author = {{Bannister}, K.~W. and {Deller}, A.~T. and {Phillips}, C. and {Macquart}, J.-P. and {Prochaska}, J.~X. and {Tejos}, N. and {Ryder}, S.~D. and {Sadler}, E.~M. and {Shannon}, R.~M. and {Simha}, S. and et al.},
        title = "{A single fast radio burst localized to a massive galaxy at cosmological distance}",
      journal = {Science},
         year = 2019,
        month = aug,
       volume = {365},
       number = {6453},
        pages = {565-570},
          doi = {10.1126/science.aaw5903},
archivePrefix = {arXiv},
       eprint = {1906.11476},
 primaryClass = {astro-ph.HE},
       adsurl = {https://scixplorer.org/abs/2019Sci...365..565B}
}

@article{Spitler_2016,
   title={A repeating fast radio burst},
   volume={531},
   ISSN={1476-4687},
   url={http://dx.doi.org/10.1038/nature17168},
   DOI={10.1038/nature17168},
   number={7593},
   journal={\nat},
   publisher={Springer Science and Business Media LLC},
   author={Spitler, L. G. and Scholz, P. and Hessels, J. W. T. and Bogdanov, S. and Brazier, A. and Camilo, F. and Chatterjee, S. and Cordes, J. M. and Crawford, F. and Deneva, J. and Ferdman, R. D. and Freire, P. C. C. and Kaspi, V. M. and Lazarus, P. and Lynch, R. and Madsen, E. C. and McLaughlin, M. A. and Patel, C. and Ransom, S. M. and Seymour, A. and Stairs, I. H. and Stappers, B. W. and van Leeuwen, J. and Zhu, W. W.},
   year={2016},
   month=Mar, pages={202–205} }

@article{Macquart_2020,
   title={A census of baryons in the Universe from localized fast radio bursts},
   volume={581},
   ISSN={1476-4687},
   url={http://dx.doi.org/10.1038/s41586-020-2300-2},
   DOI={10.1038/s41586-020-2300-2},
   number={7809},
   journal={\nat},
   publisher={Springer Science and Business Media LLC},
   author={Macquart, J.-P. and Prochaska, J. X. and McQuinn, M. and Bannister, K. W. and Bhandari, S. and Day, C. K. and Deller, A. T. and Ekers, R. D. and James, C. W. and Marnoch, L. and Osłowski, S. and Phillips, C. and Ryder, S. D. and Scott, D. R. and Shannon, R. M. and Tejos, N.},
   year={2020},
   month=May, pages={391–395} }

@ARTICLE{Walters_2018,
       author = {{Walters}, Anthony and {Weltman}, Amanda and {Gaensler}, B.~M. and {Ma}, Yin-Zhe and {Witzemann}, Amadeus},
        title = "{Future Cosmological Constraints From Fast Radio Bursts}",
      journal = {\apj},
         year = 2018,
        month = mar,
       volume = {856},
       number = {1},
          eid = {65},
        pages = {65},
          doi = {10.3847/1538-4357/aaaf6b},
archivePrefix = {arXiv},
       eprint = {1711.11277},
 primaryClass = {astro-ph.CO},
       adsurl = {https://scixplorer.org/abs/2018ApJ...856...65W}
}

@article{Ocker_2021,
doi = {10.3847/1538-4357/abeb6e},
url = {https://doi.org/10.3847/1538-4357/abeb6e},
year = {2021},
month = {apr},
publisher = {The American Astronomical Society},
volume = {911},
number = {2},
pages = {102},
author = {Ocker, Stella Koch and Cordes, James M. and Chatterjee, Shami},
title = {Constraining Galaxy Halos from the Dispersion and Scattering of Fast Radio Bursts and Pulsars},
journal = {\apj}
}

@article{Rickett1990,
  author  = {Rickett, B. J.},
  title   = {Radio Propagation Through the Turbulent Interstellar Plasma},
  journal = {\araa},
  year    = {1990},
  volume  = {28},
  pages   = {561--605},
  doi     = {10.1146/annurev.aa.28.090190.003021}
}

@article{handbook,
author = {Lorimer, D. and Kramer, M.},
year = {2004},
month = {11},
pages = {},
title = {Handbook of Pulsar Astronomy},
volume = {-1},
journal = {Handbook of pulsar astronomy, by D.R. Lorimer and M. Kramer. Cambridge observing handbooks for research astronomers, Vol. 4. Cambridge, UK: Cambridge University Press, 2004}
}

@article{Lambert_1999,
doi = {10.1086/307181},
url = {https://doi.org/10.1086/307181},
year = {1999},
month = {may},
publisher = {},
volume = {517},
number = {1},
pages = {299},
author = {Lambert, H. C. and Rickett, B. J.},
title = {On the Theory of Pulse Propagation and Two-Frequency Field Statistics in Irregular Interstellar Plasmas},
journal = {\apj}
}

@article{Masui_2015,
   title={Dense magnetized plasma associated with a fast radio burst},
   volume={528},
   ISSN={1476-4687},
   url={http://dx.doi.org/10.1038/nature15769},
   DOI={10.1038/nature15769},
   number={7583},
   journal={\nat},
   publisher={Springer Science and Business Media LLC},
   author={Masui, Kiyoshi and Lin, Hsiu-Hsien and Sievers, Jonathan and Anderson, Christopher J. and Chang, Tzu-Ching and Chen, Xuelei and Ganguly, Apratim and Jarvis, Miranda and Kuo, Cheng-Yu and Li, Yi-Chao and Liao, Yu-Wei and McLaughlin, Maura and Pen, Ue-Li and Peterson, Jeffrey B. and Roman, Alexander and Timbie, Peter T. and Voytek, Tabitha and Yadav, Jaswant K.},
   year={2015},
   month=Dec, pages={523–525} }

@article{Gajjar_2018,
   title={Highest Frequency Detection of FRB 121102 at 4–8 GHz Using the Breakthrough Listen Digital Backend at the Green Bank Telescope},
   volume={863},
   ISSN={1538-4357},
   url={http://dx.doi.org/10.3847/1538-4357/aad005},
   DOI={10.3847/1538-4357/aad005},
   number={1},
   journal={\apj},
   publisher={American Astronomical Society},
   author={Gajjar, V. and Siemion, A. P. V. and Price, D. C. and Law, C. J. and Michilli, D. and Hessels, J. W. T. and Chatterjee, S. and Archibald, A. M. and Bower, G. C. and Brinkman, C. and Burke-Spolaor, S. and Cordes, J. M. and Croft, S. and Enriquez, J. Emilio and Foster, G. and Gizani, N. and Hellbourg, G. and Isaacson, H. and Kaspi, V. M. and Lazio, T. J. W. and Lebofsky, M. and Lynch, R. S. and MacMahon, D. and McLaughlin, M. A. and Ransom, S. M. and Scholz, P. and Seymour, A. and Spitler, L. G. and Tendulkar, S. P. and Werthimer, D. and Zhang, Y. G.},
   year={2018},
   month=Aug, pages={2} }

@ARTICLE{nimmo2024magnetosphericoriginfastradio,
       author = {{Nimmo}, Kenzie and {Pleunis}, Ziggy and {Beniamini}, Paz and {Kumar}, Pawan and {Lanman}, Adam E. and {Li}, D.~Z. and {Main}, Robert and {Sammons}, Mawson W. and {Andrew}, Shion and {Bhardwaj}, Mohit and {Chatterjee}, Shami and {Curtin}, Alice P. and {Fonseca}, Emmanuel and {Gaensler}, B.~M. and {Joseph}, Ronniy C. and {Kader}, Zarif and {Kaspi}, Victoria M. and {Lazda}, Mattias and {Leung}, Calvin and {Masui}, Kiyoshi W. and {Mckinven}, Ryan and {Michilli}, Daniele and {Pandhi}, Ayush and {Pearlman}, Aaron B. and {Rafiei-Ravandi}, Masoud and {Sand}, Ketan R. and {Shin}, Kaitlyn and {Smith}, Kendrick and {Stairs}, Ingrid H.},
        title = "{Magnetospheric origin of a fast radio burst constrained using scintillation}",
      journal = {\nat},
         year = 2025,
        month = jan,
       volume = {637},
       number = {8044},
        pages = {48-51},
          doi = {10.1038/s41586-024-08297-w},
archivePrefix = {arXiv},
       eprint = {2406.11053},
 primaryClass = {astro-ph.HE},
       adsurl = {https://ui.adsabs.harvard.edu/abs/2025Natur.637...48N}
}

@ARTICLE{sammons2023twoscreenscatteringcraftfrbs,
       author = {{Sammons}, Mawson W. and {Deller}, Adam T. and {Glowacki}, Marcin and {Gourdji}, Kelly and {James}, C.~W. and {Prochaska}, J. Xavier and {Qiu}, Hao and {Scott}, Danica R. and {Shannon}, R.~M. and {Trott}, C.~M.},
        title = "{Two-screen scattering in CRAFT FRBs}",
      journal = {\mnras},
         year = 2023,
        month = nov,
       volume = {525},
       number = {4},
        pages = {5653-5668},
          doi = {10.1093/mnras/stad2631},
archivePrefix = {arXiv},
       eprint = {2305.11477},
 primaryClass = {astro-ph.HE},
       adsurl = {https://ui.adsabs.harvard.edu/abs/2023MNRAS.525.5653S}
}

@article{Petroff_2019,
   title={Fast radio bursts},
   volume={27},
   ISSN={1432-0754},
   url={http://dx.doi.org/10.1007/s00159-019-0116-6},
   DOI={10.1007/s00159-019-0116-6},
   number={1},
   journal={\aapr},
   publisher={Springer Science and Business Media LLC},
   author={Petroff, E. and Hessels, J. W. T. and Lorimer, D. R.},
   year={2019},
   month=May }

@article{Cordes_1998,
doi = {10.1086/306358},
url = {https://doi.org/10.1086/306358},
year = {1998},
month = {nov},
publisher = {},
volume = {507},
number = {2},
pages = {846},
author = {Cordes, J. M. and Rickett, B. J.},
title = {Diffractive Interstellar Scintillation Timescales and Velocities},
journal = {\apj}
}

@article{Limaye_2026,
   title={A broadband study of FRB 20240114A with the Effelsberg 100-m radio telescope},
   volume={709},
   ISSN={1432-0746},
   url={http://dx.doi.org/10.1051/0004-6361/202557626},
   DOI={10.1051/0004-6361/202557626},
   journal={\aap},
   publisher={EDP Sciences},
   author={Limaye, P. and Spitler, L. G. and Manaswini, N. and Benáček, J. and Eppel, F. and Kadler, M. and Nicotera, L. and Wongphechauxsorn, J.},
   year={2026},
   month=May, pages={A47} }

@ARTICLE{Shin2024,
  author = {{Shin}, K. and {CHIME/FRB Collaboration}},
  title = "{CHIME/FRB discovery of a new repeating fast radio burst source FRB 20240114A}",
  journal = {The Astronomer's Telegram},
  year = 2024,
  volume = {16420}
}

@article{Panda_2025,
   title={Low-frequency, Wideband Study of an Active Repeater, FRB 20240114A, with the GMRT},
   volume={989},
   ISSN={1538-4357},
   url={http://dx.doi.org/10.3847/1538-4357/adeb74},
   DOI={10.3847/1538-4357/adeb74},
   number={1},
   journal={\apj},
   publisher={American Astronomical Society},
   author={Panda, U. and Roy, J. and Bhattacharyya, S. and Dudeja, C. and Kudale, S.},
   year={2025},
   month=Aug, pages={15} }

@ARTICLE{uttarkar2026fastradioburstcyclone,
       author = {{Uttarkar}, Pavan A. and {Shannon}, Ryan M. and {Gourdji}, Kelly and {Deller}, Adam T. and {Kumar}, Pravir and {Sridhar}, Navin and {Lower}, Marcus E. and {Tuntsov}, Artem and {Kulkarni}, Atharva D. and {Ho}, Simon C.-C. and et al.},
        title = "{A fast radio burst cyclone in technicolour: evidence of plasma lensing}",
      journal = {arXiv e-prints},
         year = 2026,
        month = feb,
          eid = {arXiv:2602.16409},
        pages = {arXiv:2602.16409},
          doi = {10.48550/arXiv.2602.16409},
archivePrefix = {arXiv},
       eprint = {2602.16409},
 primaryClass = {astro-ph.HE},
       adsurl = {https://scixplorer.org/abs/2026arXiv260216409U}
}

@article{Eppel_2025,
   title={Constraints on the X-ray-to-radio fluence ratio of FRB 20240114A},
   volume={695},
   ISSN={1432-0746},
   url={http://dx.doi.org/10.1051/0004-6361/202453563},
   DOI={10.1051/0004-6361/202453563},
   journal={\aap},
   publisher={EDP Sciences},
   author={Eppel, F. and Krumpe, M. and Limaye, P. and Intrarat, N. and Wongphechauxsorn, J. and Cruces, M. and Herrmann, W. and Jankowski, F. and Jaroenjittichai, P. and Spitler, L. G. and Kadler, M.},
   year={2025},
   month=Mar, pages={L10} }

@article{Bause_2026,
   title={Searching for radio emission from radio quiet magnetars with MeerKAT},
   volume={708},
   ISSN={1432-0746},
   url={http://dx.doi.org/10.1051/0004-6361/202557296},
   DOI={10.1051/0004-6361/202557296},
   journal={\aap},
   publisher={EDP Sciences},
   author={Bause, Marlon L. and Kaur, Kamalpreet and Rammala-Zitha, Isabella and Spitler, Laura G.},
   year={2026},
   month=Apr, pages={A321} }

@ARTICLE{men2024transientxhighperformancesingle,
       author = {{Men}, Yunpeng and {Barr}, Ewan},
        title = "{TransientX: A high-performance single-pulse search package}",
      journal = {\aap},
         year = 2024,
        month = mar,
       volume = {683},
          eid = {A183},
        pages = {A183},
          doi = {10.1051/0004-6361/202348247},
archivePrefix = {arXiv},
       eprint = {2401.13834},
 primaryClass = {astro-ph.IM},
       adsurl = {https://ui.adsabs.harvard.edu/abs/2024A&A...683A.183M}
}

@PHDTHESIS{2001PhDT.......123R,
       author = {{Ransom}, Scott Mitchell},
        title = "{New search techniques for binary pulsars}",
       school = {Harvard University, Massachusetts},
         year = 2001,
        month = jan,
       adsurl = {https://ui.adsabs.harvard.edu/abs/2001PhDT.......123R}
}

@ARTICLE{tian2024detectionlocalisationhighlyactive,
       author = {{Tian}, J. and {Rajwade}, K.~M. and {Pastor-Marazuela}, I. and {Stappers}, B.~W. and {Bezuidenhout}, M.~C. and {Caleb}, M. and {Jankowski}, F. and {Barr}, E.~D. and {Kramer}, M.},
        title = "{Detection and localization of the highly active FRB 20240114A with MeerKAT}",
      journal = {\mnras},
         year = 2024,
        month = sep,
       volume = {533},
       number = {3},
        pages = {3174-3193},
          doi = {10.1093/mnras/stae2013},
archivePrefix = {arXiv},
       eprint = {2408.10988},
 primaryClass = {astro-ph.HE},
       adsurl = {https://scixplorer.org/abs/2024MNRAS.533.3174T}
}

@ARTICLE{Ocker_2026,
       author = {{Ocker}, Stella Koch and {Cordes}, James M.},
        title = "{NE2025: An Updated Electron Density Model for the Galactic Interstellar Medium}",
      journal = {\apj},
         year = 2026,
        month = may,
       volume = {1002},
       number = {1},
          eid = {3},
        pages = {3},
          doi = {10.3847/1538-4357/ae5825},
archivePrefix = {arXiv},
       eprint = {2602.11838},
 primaryClass = {astro-ph.GA},
       adsurl = {https://scixplorer.org/abs/2026ApJ..1002....3O}
}

@ARTICLE{Pradeep2025,
       author = {{Pradeep E.~T.}, Sachin and {Sprenger}, Tim and {Wucknitz}, Olaf and {Main}, Robert A. and {Spitler}, Laura G.},
        title = "{Scintillometry of fast radio bursts: Resolution effects in two-screen models}",
      journal = {\aap},
         year = 2025,
        month = aug,
       volume = {700},
          eid = {A99},
        pages = {A99},
          doi = {10.1051/0004-6361/202554202},
archivePrefix = {arXiv},
       eprint = {2505.04576},
 primaryClass = {astro-ph.GA},
       adsurl = {https://scixplorer.org/abs/2025A&A...700A..99P}
}

@ARTICLE{brown2025,
       author = {{Brown}, Grayce C. and {Sheikh}, Sofia Z. and {Cruz}, Luigi F. and {Farah}, Wael and {Gajjar}, Vishal and {Gilbertson}, Christian and {Grimaldo}, Brandon and {Lam}, Michael T. and {Marquez}, Sofia L. and {McLaughlin}, Maura and {Pollak}, Alexander W. and {Siemion}, Andrew and {Singh}, Gurmehar},
        title = "{Long-term Monitoring of Scintillation in the Pulsar J0332+5434}",
      journal = {\apj},
         year = 2025,
        month = dec,
       volume = {995},
       number = {1},
          eid = {120},
        pages = {120},
          doi = {10.3847/1538-4357/ae0fff},
       adsurl = {https://ui.adsabs.harvard.edu/abs/2025ApJ...995..120B}
}

@ARTICLE{cordes1986,
       author = {{Cordes}, J.~M.},
        title = "{Space velocities of radio pulsars from interstellar scintillations.}",
      journal = {\apj},
         year = 1986,
        month = dec,
       volume = {311},
        pages = {183-196},
          doi = {10.1086/164764},
       adsurl = {https://ui.adsabs.harvard.edu/abs/1986ApJ...311..183C}
}

@ARTICLE{cordes2010measurementmodelprecisionpulsar,
       author = {{Cordes}, J.~M. and {Shannon}, R.~M.},
        title = "{A Measurement Model for Precision Pulsar Timing}",
      journal = {arXiv e-prints},
         year = 2010,
        month = oct,
          eid = {arXiv:1010.3785},
        pages = {arXiv:1010.3785},
          doi = {10.48550/arXiv.1010.3785},
archivePrefix = {arXiv},
       eprint = {1010.3785},
 primaryClass = {astro-ph.IM},
       adsurl = {https://scixplorer.org/abs/2010arXiv1010.3785C}
}

@article{Levkov_2022,
   title={Propagation Effects in the FRB 20121102A Spectra},
   volume={925},
   ISSN={1538-4357},
   url={http://dx.doi.org/10.3847/1538-4357/ac3250},
   DOI={10.3847/1538-4357/ac3250},
   number={2},
   journal={\apj},
   publisher={American Astronomical Society},
   author={Levkov, D. G. and Panin, A. G. and Tkachev, I. I.},
   year={2022},
   month=Jan, pages={109} }

@article{Simard_2018,
   title={Predicting pulsar scintillation from refractive plasma sheets},
   volume={478},
   ISSN={1365-2966},
   url={http://dx.doi.org/10.1093/mnras/sty1140},
   DOI={10.1093/mnras/sty1140},
   number={1},
   journal={\mnras},
   publisher={Oxford University Press (OUP)},
   author={Simard, Dana and Pen, Ue-Li},
   year={2018},
   month=May, pages={983–994} }

@ARTICLE{Turner_2024,
       author = {{Turner}, Jacob E. and {Lebron Medina}, Juan G. and {Zelensky}, Zachary and {Gustavson}, Kathleen A. and {Marx}, Jeffrey and {Kothapalli}, Manvith and {Cruz Vega}, Luis D. and {Lee}, Alexander and {Figueroa}, Caryelis B. and {Reichart}, Daniel E. and et al.},
        title = "{The Pulsar Science Collaboratory: Multiepoch Scintillation Studies of Pulsars}",
      journal = {\apj},
         year = 2024,
        month = dec,
       volume = {977},
       number = {2},
          eid = {205},
        pages = {205},
          doi = {10.3847/1538-4357/ad9008},
archivePrefix = {arXiv},
       eprint = {2405.19434},
 primaryClass = {astro-ph.HE},
       adsurl = {https://scixplorer.org/abs/2024ApJ...977..205T}
}

@ARTICLE{Shin_2026,
       author = {{Shin}, Kaitlyn and {Curtin}, Alice and {Fine}, Maxwell and {Pandhi}, Ayush and {Andrew}, Shion and {Bhardwaj}, Mohit and {Chatterjee}, Shami and {Cook}, Amanda M. and {Fonseca}, Emmanuel and {Gaensler}, B.~M. and et al.},
        title = "{CHIME/FRB Discovery of the Extremely Active Fast Radio Burst Source FRB 20240114A}",
      journal = {\apj},
         year = 2026,
        month = feb,
       volume = {997},
       number = {2},
          eid = {334},
        pages = {334},
          doi = {10.3847/1538-4357/ae2fc4},
archivePrefix = {arXiv},
       eprint = {2505.13297},
 primaryClass = {astro-ph.HE},
       adsurl = {https://scixplorer.org/abs/2026ApJ...997..334S}
}

@ARTICLE{caleb2025fastradioburst3,
       author = {{Caleb}, Manisha and {Nanayakkara}, Themiya and {Stappers}, Benjamin and {Pastor-Marazuela}, In{\'e}s and {Khrykin}, Ilya S. and {Glazebrook}, Karl and {Tejos}, Nicolas and {Prochaska}, J. Xavier and {Rajwade}, Kaustubh and {Mas-Ribas}, Lluis and et al.},
        title = "{A fast radio burst from the first 3 billion years of the Universe}",
      journal = {arXiv e-prints},
         year = 2025,
        month = aug,
          eid = {arXiv:2508.01648},
        pages = {arXiv:2508.01648},
          doi = {10.48550/arXiv.2508.01648},
archivePrefix = {arXiv},
       eprint = {2508.01648},
 primaryClass = {astro-ph.HE},
       adsurl = {https://scixplorer.org/abs/2025arXiv250801648C}
}

@ARTICLE{EDDpaper,
       author = {{Barr}, Ewan D. and {Bansod}, Amit and {Behrend}, Jan and {Chen}, Weiwei and {Esser}, Niclas and {Men}, Yungpen and {Winchen}, Tobias and {Wu}, Jason and {Bach}, Uwe and {Connot}, Claus and et al.},
        title = "{The EDD Radio Astronomy Backend Framework}",
      journal = {arXiv e-prints},
         year = 2026,
        month = sep,
          eid = {arXiv:2609.11177},
        pages = {arXiv:2609.11177},
          doi = {10.48550/arXiv.2609.11177},
archivePrefix = {arXiv},
       eprint = {2609.11177},
 primaryClass = {astro-ph.IM},
       adsurl = {https://scixplorer.org/abs/2026arXiv260911177B}
}

@article{Patil_2026,
   title={A Spatial Gap in the Sky Distribution of Fast Radio Burst Detections Coinciding with Galactic Plasma Overdensities},
   volume={997},
   ISSN={2041-8213},
   url={http://dx.doi.org/10.3847/2041-8213/ae2eb3},
   DOI={10.3847/2041-8213/ae2eb3},
   number={1},
   journal={\apjl},
   publisher={American Astronomical Society},
   author={Patil, Swarali Shivraj and Main, Robert A. and Fonseca, Emmanuel and McGregor, Kyle and Gaensler, B. M. and Bhardwaj, Mohit and Brar, Charanjot and Cook, Amanda M. and Curtin, Alice P. and Eadie, Gwendolyn and Joseph, Ronniy and Kahinga, Lordrick and Kaspi, Victoria and Khan, Afrokk and Kharel, Bikash and Lanman, Adam E. and Leung, Calvin and Masui, Kiyoshi W. and Ng, Mason and Nimmo, Kenzie and Pandhi, Ayush and Pearlman, Aaron B. and Pleunis, Ziggy and Sammons, Mawson W. and Sand, Ketan R. and Scholz, Paul and Shin, Kaitlyn and Siegel, Seth R. and Smith, Kendrick},
   year={2026},
   month=Jan, pages={L5} }

@article{Chawla_2025,
   title={LOFAR constraints on the repetition and environments of CHIME FRBs},
   volume={544},
   ISSN={1365-2966},
   url={http://dx.doi.org/10.1093/mnras/staf1943},
   DOI={10.1093/mnras/staf1943},
   number={4},
   journal={\mnras},
   publisher={Oxford University Press (OUP)},
   author={Chawla, Pragya and Gopinath, Akshatha and Manaswini, Ninisha and Bassa, Cees and Hessels, Jason and Kondratiev, Vlad and Michilli, Daniele and Pleunis, Ziggy},
   year={2025},
   month=Nov, pages={4079–4097} }

@ARTICLE{connor2026dsachronoscopefastradioburst,
       author = {{Connor}, Liam and {Shin}, Kaitlyn and {Ravi}, Vikram and {Ocker}, Stella Koch and {Law}, Casey J. and {Sharma}, Kritti and {McCarty}, Samuel and {Hallinan}, Gregg and {Chatterjee}, Shami and {Cordes}, James M. and et al.},
        title = "{The DSA/Chronoscope fast radio burst survey: forecasts and science overview}",
      journal = {arXiv e-prints},
         year = 2026,
        month = aug,
          eid = {arXiv:2608.30943},
        pages = {arXiv:2608.30943},
          doi = {10.48550/arXiv.2608.30943},
archivePrefix = {arXiv},
       eprint = {2608.30943},
 primaryClass = {astro-ph.HE},
       adsurl = {https://scixplorer.org/abs/2026arXiv260830943C}
}

@article{Macquart_2013,
doi = {10.1088/0004-637X/776/2/125},
url = {https://doi.org/10.1088/0004-637X/776/2/125},
year = {2013},
month = {oct},
publisher = {The American Astronomical Society},
volume = {776},
number = {2},
pages = {125},
author = {Macquart, Jean-Pierre and Koay, Jun Yi},
title = {TEMPORAL SMEARING OF TRANSIENT RADIO SOURCES BY THE INTERGALACTIC MEDIUM},
journal = {\apj}
}

@article{Reardon_2019,
   title={Modelling annual and orbital variations in the scintillation of the relativistic binary PSR J1141−6545},
   volume={485},
   ISSN={1365-2966},
   url={http://dx.doi.org/10.1093/mnras/stz643},
   DOI={10.1093/mnras/stz643},
   number={3},
   journal={\mnras},
   publisher={Oxford University Press (OUP)},
   author={Reardon, D J and Coles, W A and Hobbs, G and Ord, S and Kerr, M and Bailes, M and Bhat, N D R and Venkatraman Krishnan, V},
   year={2019},
   month=Mar, pages={4389–4403} }

\appendix

\section{Simulation method}
\label{sec:appendix_simulation}

We assume a two-dimensional Gaussian amplitude distribution over a thin screen and sample it uniformly with random points within a disk of three standard deviations. The amplitude distribution represents the long-time ensemble average distribution while the points represent the discrete scattered images that are visible at a given time. When the two-dimensional angular position $\boldsymbol{\theta}$ of $N$ number of images is expressed in units of the standard deviation of the Gaussian amplitude distribution, the electric field can be computed as
\begin{equation}
    E(\nu) = \sum_n^N \exp\left( -\frac{1}{2}\boldsymbol{\theta}_n^2 + i\phi_n + i \frac{\nu}{\nu_s}\boldsymbol{\theta}_n^2 \right)
\end{equation}
where $\phi_n$ are uncorrelated random phases drawn uniformly from $[0,2\pi)$. In this formulation, the computation is numerically simple.

The scintillation bandwidth $\nu_s$ can be set without having to start from physical parameters like the screen's distance, which are usually unknown. The canonical notation is recovered by inserting the scintillation bandwidth in terms of the speed of light $c$ and the effective distance $D_\text{eff}$:
\begin{equation}
    \nu_s = \frac{c}{\pi D_\text{eff} \theta_s^2}
\end{equation}
where $\theta_s$ is the standard deviation of the Gaussian amplitude distribution.

\subsection{Derivation of Lorentzian ACF}

The spectral ACF is defined by the average multiplication of intensities at two frequencies $\nu_1$ and $\nu_2$ that are separated by a fixed frequency lag $\Delta \nu$ and that are each divided by the expectation value of the intensity:
\begin{equation}
    \text{ACF}(\Delta\nu) = \frac{\left\langle I(\nu_1)I(\nu_2) \right\rangle}{\langle I \rangle^2} - 1
\end{equation}
Simplifying $I=\vert E \vert^2$ because all proportionality constants cancel out, we obtain
\begin{equation}
    \langle I(\nu) \rangle = \left\langle E(\nu) \bar{E}(\nu) \right\rangle
\end{equation}
where $\bar{E}$ is the complex conjugate of $E$.

In the limit of many images, each small region in the screen contains many images, such that the random phase factors of all the images in a small region add to a complex Gaussian random variable $f$:
\begin{equation}
    f(\boldsymbol{\theta}) = \sum_m \text{e}^{i \phi_m}
\end{equation}
where $\phi_m$ are the phases of the individual images in a small region around $\boldsymbol{\theta}$. Expressed in the continuous limit, the electric field becomes
\begin{equation}
    E(\nu) = \int f(\boldsymbol{\theta})\exp\left( -\frac{1}{2}\boldsymbol{\theta}^2 + i \frac{\nu}{\nu_s}\boldsymbol{\theta}^2 \right) \, \text{d}^2 \theta
\end{equation}
Again, constant factors cancel out. Furthermore, $f$ is uncorrelated at different locations in the screen, so we can define
\begin{equation}
    \left\langle f(\boldsymbol{\theta})\bar{f}(\boldsymbol{\theta}') \right\rangle = \delta^{(2)}(\boldsymbol{\theta}-\boldsymbol{\theta}')
\end{equation}
Now we can compute the expectation value $\langle I \rangle$:
\begin{equation}
\begin{split}
    \langle I(\nu) \rangle &= \int \exp\left( -\boldsymbol{\theta}^2 \right) \, \text{d}^2 \theta \\
    &= \pi
\end{split}
\label{eq:expv_I}
\end{equation}
The main term of the ACF contains four integrations over the screen:
\begin{equation}
\begin{split}
    \left\langle I(\nu_1)I(\nu_2) \right\rangle = \left\langle E(\nu_1)\bar{E}(\nu_1)E(\nu_2)\bar{E}(\nu_2) \right\rangle
\end{split}
\end{equation}
This expression leads to an expectation value that can be separated via the complex Isserlis theorem:
\begin{equation}
\begin{split}
    \left\langle f(\boldsymbol{\theta})\bar{f}(\boldsymbol{\theta}')f(\boldsymbol{\theta''})\bar{f}(\boldsymbol{\theta}''') \right\rangle =& \left\langle f(\boldsymbol{\theta})\bar{f}(\boldsymbol{\theta}') \right\rangle\left\langle f(\boldsymbol{\theta''})\bar{f}(\boldsymbol{\theta}''') \right\rangle \\
    &+ \left\langle f(\boldsymbol{\theta})\bar{f}(\boldsymbol{\theta}''') \right\rangle\left\langle f(\boldsymbol{\theta''})\bar{f}(\boldsymbol{\theta}') \right\rangle
\end{split}
\end{equation}
The first term leads again to Eq.~\ref{eq:expv_I} while the second term mixes the frequencies. Thus, we obtain
\begin{equation}
    \text{ACF}(\Delta\nu) = \frac{1}{\pi^2}\left\vert \int \exp\left( -\boldsymbol{\theta}^2 + i \frac{\nu_1-\nu_2}{\nu_s}\boldsymbol{\theta}^2 \right) \, \text{d}^2 \theta \right\vert^2
\end{equation}
This integral can be solved by going to polar coordinates and integrating over the squared radial component:
\begin{equation}
    \pi\int_0^\infty \exp\left( -r + i \frac{\nu_1-\nu_2}{\nu_s}r \right) \, \text{d}r = \frac{\pi}{-1+i \frac{\nu_1-\nu_2}{\nu_s}}
\end{equation}
Finally, we derived
\begin{equation}
    \text{ACF}(\Delta\nu) = \frac{1}{1+\left(\frac{\Delta\nu}{\nu_s}\right)^2}
\end{equation}

The Fourier transform of the ACF is the power spectrum:
\begin{equation}
P(f) = \pi\nu_s\,\exp\!\left(-2\pi\nu_{\rm s}\,|f|\right)
\label{eq:power_spec_deriv}
\end{equation}

\end{document}